\documentclass[a4paper,12pt]{article}
\usepackage{graphicx}

\newcommand{\ds}{\displaystyle }

\title{On recurrence of random walks with long-range steps generated by fractional Laplacian matrices on regular networks and simple cubic lattices}

\author{ {\sl T.M. Michelitsch$^{1}$\footnote{Corresponding author, e-mail~: michel@lmm.jussieu.fr },
B.A. Collet$^{1}$}, A.P. Riascos$^2$ \\ \\ {\sl A. F. Nowakowski$^3$, F.C.G.A. Nicolleau$^3$}
\\ \\
$^1$ Sorbonne Universit\'es \\ Universit\'e Pierre et Marie Curie (Paris 6) \\ Institut Jean le Rond d'Alembert, CNRS UMR 7190 \\
4 place Jussieu, 75252 Paris cedex 05, France
\\ \\
$^2$ Department of Civil Engineering, Universidad Mariana\\ San Juan de Pasto, Colombia
\\ \\
$^3$ Sheffield Fluid Mechanics Group\\
Department of Mechanical Engineering\\
University of Sheffield\\
Mappin Street, Sheffield S1 3JD,
United Kingdom
\\
}
\begin{document}

\maketitle

\newpage

\begin{abstract}
We analyze a random walk strategy on undirected regular networks involving power matrix functions of the type $L^{\frac{\alpha}{2}}$ where $L$ indicates a `simple' Laplacian matrix.
We refer such walks to as `Fractional Random Walks' with admissible interval $0<\alpha \leq 2$.
We deduce for the Fractional Random Walk probability generating functions (network Green's functions).
From these analytical results
we establish a generalization of Polya's recurrence theorem for Fractional Random Walks on $d$-dimensional infinite lattices:
The Fractional Random Walk is transient for dimensions $d > \alpha$ (recurrent for $d\leq\alpha$) of the lattice.
As a consequence for $0<\alpha< 1$ the Fractional Random Walk is
transient for all lattice dimensions $d=1,2,..$ and in the range $1\leq\alpha < 2$ for dimensions $d\geq 2$. Finally, for $\alpha=2$
Polya's classical recurrence theorem is recovered, namely
the walk is transient only for lattice dimensions $d\geq 3$.
The generalization of Polya's recurrence theorem remains valid for the class of random walks with L\'evy flight asymptotics for long-range steps.
We also analyze for the Fractional Random Walk
mean first passage probabilities, mean residence times, mean first passage times, and global mean first passage times (Kemeny constant).
For the infinite 1D lattice (infinite ring) we obtain for
the transient regime $0<\alpha<1$ closed form
expressions for the fractional lattice Green's function matrix containing the escape and ever passage probabilities. The ever passage probabilities fulfill
Riesz potential power law decay asymptotic behavior for nodes far from the departure node.
The non-locality of the Fractional Random Walk is generated by the non-diagonality
of the fractional Laplacian matrix with L\'evy type heavy tailed inverse power law decay for the probability of long-range moves.
This non-local and asymptotic behavior of the Fractional Random Walk introduces small world properties with emergence of L\'evy flights on large (infinite) lattices.
\end{abstract}


\section{Introduction}

Due to the rapid growth of online networks and search engines such as for instance google, there is an increasing interest in improved and faster search and navigation
strategies on complex networks \cite{newmann,albert2002,NohRieger,gonzales,rathkiewics,riascos12}.
The number of systems which can be conceived as complex networks indeed is huge. They include biological, social-, friendship networks, human cities, electricity networks,
the water supply networks, transport networks (rivers, streets), computer networks such as the world wide web but also crystalline
structures of solids and numerous further examples can be denominated. Due to this huge variety of different fields and contexts,
the study of dynamic processes on networks has become a vast interdisciplinary field.
On the other hand many of these processes such as internet search, the spread of rumors, news headlines, propagation of pandemic deceases, foraging, the transitions in chemical reactions and
many other examples in very different contexts can be considered as random walks on abstract sets of points (nodes) on networks (graphs).

It was Polya in 1921
who was probably one of the first to furnish a thorough analysis of Markovian time-discrete random walks on periodic $d$-dimensional lattices. In these `Polya walks' the walker
is allowed to step with
equal probability only to any of its neighbor nodes \cite{polya,montroll,montroll-weiss,hudges}. Polya proved for this kind of random walk that the walker is sure to
return to its starting node for dimensions $d=1,2$ of the lattice whereas for dimensions $d>2$ a finite escape probability (probability of never return) exists.
This celebrated result has become known as {\it Polya theorem} or {\it Recurrence Theorem} \cite{polya,hudges}. As one of the subjects of the present paper we will generalize this theorem to
`Fractional Random Walks' (FRWs).

Noh and Rieger \cite{NohRieger} considered important characteristics of `Normal Random Walks' (NRWs) in complex networks which are a generalization of the Polya type walk to networks of variable degree of the nodes. In that paper
were deduced characteristics
such as mean first passage times (MFPTs) and first passage probabilities.
Watts and Strogatz
\cite{Watts} showed that in many `real world networks' features like small world emerge that are not captured by the previously mentioned classical network models \cite{dorogotsev}.
Indeed in the meantime numerous further models were developed to generate new types of sophisticated small-world networks among them new features as randomly generated such as the Er\"os-R\'enyi network \cite{Erdos}, scale-free (self-similar) and fractal networks \cite{albert2002,dorogotsev}.

A generalization of the NRW concept \cite{NohRieger} to L\'evy type random walks on complex undirected networks was presented by Riascos and Mateos \cite{riascos12}.
They demonstrated that if L\'evy type navigation strategy is performed on large world network, small world properties are emerging increasing efficiency compared to as a NRW.

The fractional calculus approach has turned out to be a powerful analytical tool to describe a large variety of phenomena such as anomalous diffusion and fractional transport \cite{metzler1,klafter} (and the references therein).
Usually the fractional approach is often applied to problems in continuous spaces. A fractional lattice approach has been suggested by Tarasov \cite{tarasov}.

In many practical applications in the development of search strategies the simple question occurs how the average number of necessary steps can be reduced until
a result is found. In the random walk picture this question corresponds to find random walk strategies reducing first passage times. One goal of this paper is to
demonstrate that this can be achieved for random walk strategies based on fractional Laplacian matrices (FRWs).

The present paper is arranged as follows.
In the subsequent section \ref{sec1} we invoke some basic general features on Markovian random walks on undirected regular networks. We invoke some important spectral properties of the transition matrix for the FRW
and determine probability generating functions (which we refer alternatively to as Green's functions) for the node occupation probabilities and
first passage probabilities which are highly powerful analytical tools in the determination of these probabilities.
To keep our demonstration as simple as possible we especially focus on regular networks,
i.e. networks having constant degree for all nodes such as $d$-dimensional cubic primitive periodic and infinite lattices ($d=1,2,3,4,...$).
Nevertheless, many of the obtained results can be generalized to more complex network types.

In section \ref{sec2} we deduce the Green's functions of $d$-dimensional lattices and analyze
infinite lattice limits for the probabilities that a node is ever visited including probabilities of ever return.
In this way we establish a
generalization of Polya's recurrence theorem for Fractional Random Walks holding in the limit of infinite lattices.
Further,
we also obtain for $d$-dimensional periodic and infinite lattices exact expressions for
the mean first passage times (MFPT) and global MFPT (Kemeny constant) in terms of spectral properties of the fractional Laplacian matrix.

In section \ref{explcit1Dring} we develop for the transient regime of the Fractional Random Walk explicit
expressions for the 1D infinite lattice (infinite ring) for the ever passage probabilities and escape probabilities
and obtain power law evanescent asympotic behavior of Riesz potential forms for the ever passage probabilities for nodes far from the departure node.

The main dynamic effect is that a FRW performed on a large world network appears as a walk in a small world network.
The long-range moves appearing in the FRW make the dynamics of the FRW remarkably rich.
The present study is aiming to demonstrate some of these dynamic effects and complement some previous studies on the subject
\cite{riascos12,riascos-fracdyn,riascos-fracdiff,michelitsch-riascos2017,michelJphysA,michelchaos,michelchaos2}.

\section{Basic notions on Markovian time discrete random walks}
\label{sec1}

\subsection{The fractional Laplacian matrix and the FRW on regular undirected networks}

We analyse random walks on regular undirected connected networks (graphs) of $N$ nodes which we denote by $p=0,..,N-1$.
In regular networks considered in the present paper all nodes (vertices) $p$ have constant degree $K_p=K$ $\forall p=0,..,N-1$.
This is the case in lattice structures that are considered in the present paper.
Whether or not a pair of nodes $p,q$ is connected (by an edge) is described by the $N\times N$ adjacency matrix $A$ with elements $A_{pq}=1$ if the nodes $p,q$
are connected and $A_{pq}=0$ otherwise. Further we assume $A_{pp}=0$.
In an undirected network the connections (edges) between nodes have no direction and as a consequence
the adjacency matrix is symmetric $A_{pq}=A_{qp}$. The properties of the network are characterized by the Laplacian matrix which we can write
in its spectral representation \cite{newmann,miegem}
\begin{equation}
 \label{Laplacianmatrix}
 L_{pq} = \delta_{pq}K_p -A_{pq} = \sum_{j=1}^N\mu_j\langle p |\Psi_j\rangle\langle\Psi_j|q\rangle
\end{equation}
where we adopt in this paper the common Dirac's bra-ket notation and we have the symmetry $L_{pq}=L_{qp}$ of the Laplacian matrix of undirected networks.
In the present paper we assume constant degree $K_p=K$ for any node $p$ thus the Laplacian matrix takes the form
$L=K{\hat 1}-A$ where ${\hat 1}$ denotes the $N\times N$ unity matrix.
Due to the symmetry of the Laplacian matrix the set of eigenvectors constitute a complete $N$-dimensional ortho-normal canonic basis\footnote{$\langle\Psi_i|\Psi_j\rangle = \delta_{ij}$ and
$\sum_{m=1}^N\langle i|\Psi_m\rangle\langle\Psi_m|j\rangle = \delta_{ij}$}.
The degree $K_p$ of a node $p$ counts the number of connections of node $p$ with other nodes. This is expressed by the relationship
$K_p=\sum_{q=0}^{N-1}A_{pq}$.
It follows that the constant vector which we denote as
$|\Psi_1\rangle=\frac{1}{\sqrt{N}}(1,..,1)$ is eigenvector of the Laplacian matrix $L$ to the zero eigenvalue $\mu_1=0$.
Generally the Laplacian matrix is positive-semidefinite and
in connected networks the vanishing eigenvalue $\mu_1$ appears uniquely together with $N-1$ positive eigenvalues $\mu_1=0 < \mu_2 \leq..,\leq \mu_N$ \cite{miegem}.

To analyze random walks on networks, we introduce the probability vector $\vec{P}_t = \sum_{p=0}^{N-1} P_t(p)|p\rangle $ having the occupation probabilities $P_t(p)$ of the nodes $p$
as Cartesian components where variable $t$ denotes the time.
As the random walker is for sure somewhere on the network, the normalization condition
$\sum_{q=0}^{N-1} P_t(p) =1$ is fulfilled for the entire time of observation $0\leq t < \infty$ where we define $t=0$ as the time of departure of the random walker.
We consider time-discrete random walks at integer times $t=0,1,2,..$  where the walk is assumed to start at $t=0$
and during one time increment
$\delta t=1$ the random walker is allowed to move from one to another node where only steps between connected nodes ($A_{pq}\neq 0$) are allowed.

The time evolution of the occupation probabilities for a Markovian walk is governed by a discrete master equation where we utilize alternatively matrix and index notations \cite{hudges}
\begin{equation}
 \label{timemaster}
 P_{t+1}(p) = \sum_{q=0}^{N-1} {\cal W}_{pq} P_t(q) ,\hspace{2cm} \vec{P}_{t+1} =  {\cal W}\cdot \vec{P}_t .
\end{equation}
The $N\times N$ matrix ${\cal W}=:W(\delta t=1)$ is referred to as {\it transition matrix}
connecting the probabilities $\vec{P}_{t+1}$ with $\vec{P}_{t}$.
Despite we consider in the present analysis time-discrete random walks, a transition to time continuous walks is straight-forward \cite{michelitsch-riascos2017}
\footnote{Letting $\delta t \rightarrow 0$ infinitesimal yields the transition matrix of
time-continuous random walks.}.
For random walks taking place as Markovian processes the transition matrix of one time step
${\cal W}=W(\delta t)$ is constant depending only on the time step $\delta t$, but not on the history of the walk.
The time-evolution (\ref{timemaster}) of the transition matrix iterating $t=n$ time-steps then is

\begin{equation}
\label{timeevol}
P_n(p) =\sum_{p=0}^{N-1}\langle p  |{\cal W}^n q \rangle P_0(q)
\end{equation}
where $W^0={\hat 1}$ denotes the $N\times N$ unity matrix and where subsequently we utilize synonymously $\langle p  |{\cal W}^t q \rangle= W_{pq}(t)$ for the elements of the transition matrix
${\cal W}^t=W(t)$. The transition matrix fulfills the normalization condition
\begin{equation}
\label{norma1}
 \sum_{p=0}^{N-1} W_{pq}(t) =1 ,\hspace{2cm} 0 \leq W_{pq}(t) \leq 1
\end{equation}
reflecting the normalization of the occupation probabilities.
We also have the restriction $0 \leq W_{pq}(t)\leq 1$ allowing the probability interpretation to be maintained for the entire observation time $0 \leq t=n < \infty$.
As mentioned we confine us here on undirected regular networks which are characterized by the symmetry property $W_{pq}(t)=W_{qp}(t)$ reflecting the fact that there is no
preference for moves between $p$ to $q$ and vice versa.
For the analysis to follow it is worthy to consider the spectral
properties of the transition matrix. As the transition matrix is symmetric (self-adjoint), it can be expressed by its (purely real) eigenvalues $\lambda_m$ and
eigenvectors $|\Psi_m\rangle$ as

\begin{equation}
\label{spectral}
 {\cal W} = W(t=1)= \sum_{m=1}^N \lambda_m|\Psi_m\rangle\langle\Psi_m| ,\hspace{1cm}  W(t=n)_{pq} = ({\cal W}^n)_{pq} = \sum_{m=1}^N (\lambda_m)^n
 \langle p|\Psi_m\rangle\langle\Psi_m|q\rangle .
\end{equation}

In connected ergodic networks with constant degree the stationary distribution is constituted by the equal-distribution
$W_{pq}(t\rightarrow \infty) = \langle p|\Psi_1\rangle\langle\Psi_1|q\rangle =\frac{1}{N}$ ($\forall p,q =0,..,N-1$) \cite{miegem,riascos12}
\footnote{From $\lim_{n\rightarrow\infty }{\cal W}^n =W(\infty)= |\Psi_1\rangle\langle\Psi_1|$
follows that $\lambda_1=1$ is unique where the remaining $N-1$ eigenvalues $|\lambda_m|<1$ $m=2,..,N$ where all eigenvalues are real due to ${\cal W}_{pq}={\cal W}_{qp}$.}.

Now we relate network properties with the random walk dynamics by means of the Laplacian matrix. For the NRW on a regular network
the transition matrix of one time step takes the following form \cite{NohRieger,riascos12,michelitsch-riascos2017}

\begin{equation}
 \label{changesite}
 {\cal W}_{pq}= \delta_{pq} -\frac{1}{K_p}L_{pq} = \frac{1}{K}A_{pq} ,\hspace{1cm} K=\frac{1}{N}tr(L)
\end{equation}
where $tr(..)$ denotes the trace $\sum_{p=0}^{N-1}(..)_{pp}$ of a matrix and $\frac{1}{K_p}L_{pq}=\frac{1}{K}L_{pq}$ the (for regular networks considered in the present paper symmetric) generator matrix
of the random walk.
We notice that ${\cal W}_{pp}=0$, that is the walker is forced to change its node at each time step, moving with equal probability $\frac{1}{K}$
to any next neighbor node\footnote{We identify `next neighbor node' with `connected node'.}. The transition matrix
(\ref{changesite}) is invariant by rescaling $L \rightarrow \zeta L$ of the Laplacian matrix by any non-zero
scaling factor $\zeta$\footnote{This scaling invariance is absent in the time continuous random walks when the transition matrix is defined as in \cite{michelitsch-riascos2017}.}.
We utilize this simple observation subsequently for the physical interpretation of the dynamics of the fractional random walk.

It follows from (\ref{changesite}) that the eigenvalues of ${\cal W}$ and of the Laplacian
matrix $L$ are related by $\lambda_m=1-\frac{\mu_m}{K}$ ($m=1,2,..,N$) where both the transition
matrix and the Laplacian matrix as well as any matrix functions of these matrices have an identical
space of eigenvectors with the canonic set $|\Psi_j\rangle$ ($j=1,2,..,N$).

Transition matrix (\ref{changesite}) defines a Normal Random Walk (NRW) on regular networks with constant degree.
After $n=0,1,2,..$ time steps the transition matrix elements take the form

\begin{equation}
\label{timevol}
W_{pq}(n) = P_n(p,q)= [{\cal W}^n]_{pq}= \frac{1}{K^n} Z_n(p,q)
\end{equation}
indicating the probability that node $p$ is occupied by the walker at the $n^{th}$-time steps walk departing at node $q$.
In this relation we have introduced
\begin{equation}
\label{numberpath}
\begin{array}{l}
 Z_1(p,q)= A_{pq} , \hspace{1cm} n=1 \\ \\
Z_n(p,q)= (A^n)_{pq} = \sum_{j_1,j_2,..,j_{n-1}}A_{pj_{n-1}}A_{j_{n-1}j_{n-2} }..A_{j_1q} ,\hspace{1cm} n=2,3,..
\end{array}
\end{equation}
indicating the number of possible paths the walker can take when performing a walk of $n$ time steps starting at node $q$ ending at node $p$.
Each of these paths occurs in (\ref{timevol}) with equal probability $\frac{1}{K^n}$ where $K^n$ indicates the number of
paths a random walker can chose performing $n$ time steps where all paths depart from the same node.

Random walks of the type (\ref{timevol}), (\ref{numberpath}) with equal probability
of any possible path, we refer to as Polya walks \cite{polya,hudges} where for regular networks considered in the present paper the notions 'Polya walk' and
'Normal Random Walk (NRW)' can be synonymously used.
The equal probability distribution of paths is characteristic
for Polya type walks and
is not any more true for FRWs subsequently analyzed.
Summarizing (\ref{numberpath}) over all possible paths starting at node $q$ of $n$ time steps, we obtain
the total number of $n$-time step paths\footnote{Where the summation over all paths indeed is a discrete network version of Feynman's path integral.}
$\sum_{p=0}^{N-1}Z_n(p,q) =  \sum_{p=0}^{N-1} K^n W_{pq}(n) =K^n$ starting at $q$
reflecting the normalization condition of the transition matrix $W_{pq}(n)$ at any time step.

Main subject of the present analysis is to study a generalization of the NRW which we refer to as {\it Fractional Random Walk (FRW)} where
the Laplacian matrix (\ref{Laplacianmatrix}) which was employed in (\ref{changesite})
is replaced by a fractional non-integer power matrix function of the form

\begin{equation}
\label{fracti}
L^{\frac{\alpha}{2}} = \sum_{m=2}^N (\mu_m)^{\frac{\alpha}{2}} |\Psi_m\rangle\langle\Psi_m| ,\hspace{1cm} 0<\alpha \leq 2
\end{equation}
which we refer to as {\it Fractional Laplacian matrix}.
It is important to notice that the admissible interval for $\alpha$ in order to maintain the good properties of a random walk generator matrix is
$0 < \alpha \leq 2 $ \cite{michelitsch-riascos2017}. For $\alpha=2$ (\ref{fracti}) recovers the Laplacian matrix (\ref{Laplacianmatrix}) where the FRW then recovers the NRW (Polya walk).
We define the transition matrix of one time step for the FRW corresponding to (\ref{changesite}) \cite{riascos-fracdiff,riascos-fracdyn}

\begin{equation}
\label{FRWtrans}
{\cal W}^{(\alpha)}_{pq} = W^{(\alpha)}(t=1)= \delta_{pq} -\frac{1}{K^{(\alpha)}}(L^{\frac{\alpha}{2}})_{pq}
=: \frac{1}{K^{(\alpha)}}(A^{(\alpha)})_{pq} ,\hspace{1cm} 0 < \alpha \leq 2 .
\end{equation}
Here we introduced the fractional degree $K^{(\alpha)}$ which is a constant in regular networks and given by the constant diagonal element of the fractional Laplacian matrix

\begin{equation}
 \label{fractionaldegree}
 K^{(\alpha)} = [L^{\frac{\alpha}{2}}]_{pp}= \frac{1}{N}tr(L^{\frac{\alpha}{2}}) = \frac{1}{N} \sum_{m=1}^N(\mu_m)^{\frac{\alpha}{2}} ,\hspace{1cm} 0 < \alpha \leq 2 .
\end{equation}
So we observe that the diagonal element of the transition matrix ${\cal W}^{(\alpha)}_{pp}= \frac{1}{N}tr({\cal W}^{(\alpha)}) =0$
is vanishing as for the NRW and further due to $\mu_1=\mu_1^{\frac{\alpha}{2}}= 0$ the stationary distribution (ergodicity) for the FRW
$\langle p|\Psi_1\rangle\langle\Psi_1|q\rangle = \frac{1}{N}$ together with $\lambda_1=1$ of the fractional transition matrix is maintained.
We introduced in relation (\ref{FRWtrans}) the fractional adjacency matrix

\begin{equation}
 \label{fractionaldjacncy}
 A^{(\alpha)}_{pq} = \delta_{pq}K^{(\alpha)}- (L^{\frac{\alpha}{2}})_{pq}  \geq 0 ,\hspace{1cm} 0 < \alpha \leq 2
\end{equation}
where we have analogous properties as in the non-fractional case $K^{(\alpha)} = \sum_{q=0}^{N-1} A^{(\alpha)}_{pq}$ reflecting conservation of eigenvalue zero and
corresponding eigenvector $|\Psi_1\rangle$ of the fractional Laplacian matrix. The fractional adjacency matrix $A^{(\alpha)}_{pq}$
has uniquely non-negative off diagonal elements $A^{(\alpha)}_{pq} = - (L^{\frac{\alpha}{2}})_{pq} \geq 0$ ($p\neq q$)\footnote{where it was demonstrated \cite{michelitsch-riascos2017} that $(L^{\frac{\alpha}{2}})_{pq}$ has
non-positive off-diagonal elements for $0<\alpha\leq 2$.}
and zero diagonal elements $A^{(\alpha)}_{pp}=0$
allowing the probability interpretation of (\ref{FRWtrans}) fulfilling the necessary conditions (i) $0 \leq W^{(\alpha)}_{pq}(t) \leq 1$ and (ii)
$\sum_{q=0}^{N-1}W^{(\alpha)}_{pq}(t) =1$. As already mentioned for regular undirected networks
these good properties of the fractional Laplacian matrix allowing definitions (\ref{fractionaldegree})
and (\ref{fractionaldjacncy}) with (i) and (ii) are fulfilled within  $0 < \alpha \leq 2$. We notice that this range of exponent is exactly the range of definition of L\'evy
index occuring in the context of L\'evy flights and
L\'evy ($\alpha$-stable) distributions. Moreover, the emergence of L\'evy flights for FRWs on $d$-dimensional lattices
has been demonstrated recently \cite{riascos-fracdyn,riascos-fracdiff,michelitsch-riascos2017,michelJphysA}. A short demonstration is given in appendix \ref{appendb}.

\subsection{Some general remarks on first passage probabilities and mean first passage times}
\label{passage}

In this section we evoke some basic relations between occupation probabilities and first passage probabilities and
their statistical interpretations as far as required for the present analysis of the FRW.
The efficiency of a random walk strategy to explore the network can be measured by first passage quantities such as mean first passage probability and mean first passage times (MFPT) for a node.
In the deductions to follow we first consider finite networks, i.e. with a finite number $N$ of nodes and analyze the limit $N\rightarrow\infty$ of infinite networks.
It turns out that new features such as transience of the random walk may emerge in the limiting cases of infinite networks. To this end we evoke some general relations on first passage events
as far as they are needed for the subsequent analysis.
For a thorough analysis and further discussions of general properties we refer to  \cite{hudges,miegem,doyle,kemeny,Feller}.

In the subsequent analysis when a quantity $B$ refers to the Fractional random Walk (FRW), we employ the notation $B^{(\alpha)}$ with a superscript $(..)^{(\alpha)}$.
Otherwise for general relations as well as for NRWs we utilize $B$. The following definitions and notions will be used.
\newline\newline\noindent
1) $P_n(p,q)$ denotes the occupation probability of node $p$ by a random walker starting at node $q$ undertaking a walk of $n$ time steps. This probability coincides with the ratio
of the number $Z_n(p,q)$ of paths starting at $q$ ending at $p$ of $n$ time steps and the total number $K^n$ of paths of $n$ time steps with the same departure node.
The occupation probabilities $P_n(p,q)$ were already
defined in above relation (\ref{timevol}).
\newline\newline\noindent
2) As a generalization of (\ref{numberpath}) we interpret
$(K^{(\alpha)})^n$ (where $K^{(\alpha)}$ denotes the fractional degree (\ref{fractionaldegree}))
as the `fractional' number of allowed paths for a FRW of $t=n$ time steps with the same starting node.
\newline\newline \noindent
3) The quantity
$Z^{(\alpha)}_n(p,q)= ((A^{(\alpha)})^n)_{pq} = \sum_{j_1,..,j_{n-1}}A^{(\alpha)}_{pj_{n-1}} ..A^{(\alpha)}_{j_1q}$ indicates the `fractional' number of
paths of $n$ time steps starting at node $q$ ending at node $p$ where it turns out that the equal distribution of the Polya walk
is not true for the FRW when $0<\alpha<2$.
The occupation probability of the FRW is $P_n^{(\alpha)}(p,q) =W^{(\alpha)}_{pq}(n)= \frac{Z^{(\alpha)}_n(p,q)}{(K^{(\alpha)})^n}$ constituting the matrix elements of the fractional
transition matrix $W^{\alpha}_{pq}(n)=\langle p ({\cal W}^{(\alpha)})^n|q\rangle $.
In the fractional case $0<\alpha<2$ the $Z^{(\alpha)}_n(p,q)$ and $(K^{(\alpha)})^n$ generally are non-negative non-integers. For $\alpha=2$ all characteristics of the Polya walk are recovered.
\newline\newline
\noindent 4)
$F_n(p,q)$ denominates the {\it first passage probability}, that is the probability that a random walker starting from node $q$ visits node $p$ at the $n^{th}$ time step for the {\it first time}.
For interpretation purposes we introduce the number $f_n(p,q)$ of first passage paths of $n$ time steps. A first passage path is a path starting at node $q$
containing node $p$ only once as end node. These are the paths of $n$ time step walks departing at node $q$
passing at node $p$ for the {\it first time}.
When $p=q$ these paths constitute closed cycles representing paths of first return to the starting node $q$.
It follows that the first passage probabilities can be represented as
$F_n(p,q)= \frac{f_n(p,q)}{K^n}$ where $K^n$ denotes the total number of possible paths of $n$ time steps starting all from node $q$ where $K$ indicates the constant degree.
In a regular undirected network the probabilities of first return $F_n(q,q)=F_n(0,0)$
are constant for all nodes and the probabilities of first passage $F_n(p,q)=F_n(q,p)$ as well as the occupation probabilities (transition matrices) represent symmetric matrices with respect
of interchanging starting and end nodes.
\newline\newline
With these definitions we can establish a relationship between the first passage probabilities $F_t$ and the occupation probabilities $P_t$ which holds for Markovian walks
\cite{NohRieger,riascos12,hudges}

\begin{equation}
\label{occupation}
P_t(p,q) = \delta_{t0}\delta_{pq} +\sum_{k=0}^tF_{t-k}(p,q)P_k(0|0)
\end{equation}
with $P_0(p,q)=\delta_{pq}$ where $F_0(p,q)=0$ and $F_1(p,q)=P_1(p,q)$ as at $t=1$ only next neighbor nodes can be visited for the first time.
Relation (\ref{occupation}) can be interpreted as follows by multiplying relation (\ref{occupation}) by the total number of possible paths $K^n$ of $n$ time steps.
In Markovian random walks the number of possible paths $Z_n(p,q)$ connecting the nodes $q$ and $p$ of $n$ time steps
can be decomposed into $Z_n(p,q) = \sum_{k=0}^nf_{n-k}(p,q)Z_k(0,0)$ ($f_0=0$). That is the number $f_{n-k}(p,q)$ of first passage paths of $n-k$ time steps
multiplied with the number $Z_k(0,0)$ of return cycles of $k$ time steps whereas all combinations $k=0,..,n$ occur as a sum reflecting the property
that first passage events at different times are exclusive events representing different paths.
Whereas the occupation probabilities are determined by (\ref{timeevol}), the {\it first passage probabilities} are uniquely determined by (\ref{occupation}).

For the determination of the probabilities $Q_n=(P_n,F_n)$ and of further characteristics
it is convenient to employ the method of probability
generating functions. The probability generating function of the probabilities $\{ Q_n \}$ is
defined as a power series having these probabilities as non-negative coefficients

\begin{equation}
 \label{probabgen}
 Q(p,q,\xi)= \sum_{n=0}^{\infty} Q_n(p,q)\xi^n \hspace{1cm} |\xi| < 1
\end{equation}
having according to Abel's theorem (at least) the radius of convergence $\xi=1$.
The $Q(p,q,\xi)=(F(p,q,\xi),P(p,q,\xi))$ stand in the following analysis for the generating functions of the first passage- and occupation probabilities, respectively.
The occupation probability generating function $P(p,q,\xi)$ is also referred to as network Green's function \cite{montroll,montroll-weiss,hudges}.

The elements of the network Green's function (at $\xi=1$) $P(p,q,\xi=1)=\sum_{t=0}^{\infty}P_t(p,q)$ indicate the mean residence times (MRT), i.e. the average number of time steps the walker occupies a node $p$ (when starting the walk at node $q$) for an infinite time of observation $t\rightarrow \infty$. For discrete time random walks defined as in (\ref{changesite}) where the walker at any time step moves to another node (as $W_{pp}=0$), the MRT $P(p,q,1)$ counts the average number of visits of a node
during an infinite observation time. It follows that a walk is {\it transient} if the average number of visits of a node is finite
$P(p,q,\xi=1) <\infty$, and further a walk is {\it recurrent} if a node $p$ infinitely often is visited $P(p,q,\xi=1)\rightarrow \infty$.
An instructive study of the MRT and MFPT among other features of L\'evy flights versus L\'evy walks in the 1D continuous space is given in \cite{barkai}, and a further analysis of first escapes and arrivals in finite domain is
performed in \cite{Dybiec-Nowac}.

The zero order $F_0(p,q)$ in the series for $F(p,q,\xi)$ is vanishing whereas $P_0(p,q)=\delta_{pq}$.
The probability generating function (\ref{probabgen})
can be read as discrete Laplace transform  by $\xi=e^{-s}$ converging for $\Re(s) >0$.
We mention this point as technically for the determination of the moments the Laplace transform is more convenient to use
$\langle (t^m)_{pq} \rangle = (-1)^m\frac{d^m}{ds^m}Q(p,q,e^{-s})|_{s=0}$.
(\ref{occupation}) can be identified $n^{th}$ orders $\sim \xi^n$ ($n=1,2,..$) of the functional identity

\begin{equation}
\label{genera1}
P(p,q,\xi)-\delta_{pq} = F(p,q,\xi)P(0,0,\xi) .
\end{equation}
Thus the generating function for the first passage probabilities is obtained as

\begin{equation}
 \label{firstpassagepro}
 F(p,q,\xi) = \frac{P(p,q,\xi)-\delta_{pq}}{P(0,0,\xi)} ,\hspace{1cm} F(\xi)= \frac{1}{P(0,0,\xi)}\left(P(\xi)-{\hat 1}\right)
\end{equation}
where in the second equation we write this relation in matrix form.
For our subsequent analysis of the FRW it is useful to relate
(\ref{firstpassagepro}) with the spectral properties of transition matrix and (fractional) Laplacian matrix.
To this end we evaluate the generating matrix for a finite network of $N$ nodes

\begin{equation}
\label{pro}
P(\xi) = \sum_{n=0}^{\infty}{\cal W}^{n}\xi^n = [{\hat 1}-\xi {\cal W}]^{-1} =
\frac{|\Psi_1\rangle\langle\Psi_1|}{(1-\xi)} + \sum_{m=2}^{N}|\Psi_m\rangle\langle\Psi_m|\frac{1}{(1-\lambda_m\xi)} ,\hspace{0.5cm} |\xi| < 1 .
\end{equation}
The diagonal element of (\ref{pro}) which is identical for all nodes is obtained as
\begin{equation}
 \label{diagonalelement}
P(p,p,\xi)=P(0,0,\xi) = \frac{1}{N}tr(P(\xi)) = \frac{1}{N}\left(\frac{1}{(1-\xi)}+\sum_{m=2}^{N}\frac{1}{(1-\lambda_m\xi)}\right) .
\end{equation}
For $p=q$ (\ref{firstpassagepro}) contains the probabilities of first return to the starting node $F(0,0,\xi)= 1-\frac{1}{P(0,0,\xi)}$ being identical for all departure nodes.

We mention the important property that $F(p,q,\xi\rightarrow 1) = \sum_{n=1}^{\infty} F_n(p,q)$
yields the probability that
the random walker starting at node $q$ {\it ever} visits node $p$, or equivalently that the random walker visits node $p$ {\it at least once} during the infinite observation time $t \rightarrow \infty$
\cite{hudges}.
This information is hence contained in (\ref{firstpassagepro}) in the limiting case $\xi\rightarrow 1-0$. This quantity indeed is of great importance in many contexts
such as survival time models and the subsequently analyzed question of recurrence (transience) of a random walk.
For an infinite network $N\rightarrow\infty$ the contribution of the stationary distribution $\langle p|\Psi_1\rangle\langle\Psi_1|q\rangle = \frac{1}{N} \rightarrow 0$
is suppressed reflecting the property that the Green's function (\ref{pro}) in the limit $N\rightarrow \infty$ becomes a `generalized function' in the distributional sense \cite{gelfand}.
That is the matrix elements coincide
at any entry $(pq)$ whereas the suppressed stationary distribution $\langle p|\Psi_1\rangle \langle q|\Psi_1\rangle$
becomes important only when performing the infinite spectral sum
(in the sense of below defined limiting integral (\ref{infiniteGreen})),
namely we have $r_{pq}(\xi) = P_{N\rightarrow \infty}(p,q,\xi)$ as distributional identity, however, $\sum_{p=0}^{\infty}(P_{\infty}(p,q,\xi)-r_{pq}(\xi)) =1$
due to the normalization of the probabilities.

To evaluate the infinite network Green's function which we denote subsequently as
$P_{N\rightarrow \infty}(p,q,\xi) =r_{pq}(\xi)$,
its spectral representation in infinite networks is determined by the spectral sum (\ref{pro}) accounting only for the relaxing eigenvalues $|\lambda_m|<1$,
namely
\begin{equation}
 \label{infiniteGreen}
 P_{\infty}(\xi) = r(\xi) = \sum_{m=2}^{\infty}|\Psi_m\rangle\langle\Psi_m|\frac{1}{(1-\lambda_m\xi)} = \int_{\lambda_{min}}^1 \frac{|{\tilde \Psi}(\lambda)\rangle\langle{\tilde \Psi}(\lambda)|D(\lambda)}{(1-\xi\lambda)} {\rm d}\lambda
\end{equation}
where as mentioned for $N\rightarrow\infty$ the stationary contribution corresponding to $\lambda=1$ is suppressed.
Whether or not (\ref{infiniteGreen}) converges for $\xi=1$ depends on the properties of the infinite
network. On the right hand side of (\ref{infiniteGreen}) we accounted for property that
the spectra  $\lambda_m\rightarrow \lambda$ becomes continuous when $N\rightarrow\infty$ and the eigenvalue density $D(\lambda){\rm d}\lambda$ counts the number of eigenvalues within
$[\lambda,\lambda+{\rm d}\lambda]$ and can be represented\footnote{Where $\lim_{\epsilon\rightarrow 0+} \frac{1}{\pi}\Im \frac{1}{(x-i\epsilon)} = \frac{1}{\pi}\Im \frac{1}{(x-i0)} =\delta(x)$ and $\Im(..)$
indicates the imaginary part of $(..)$ with ${\tilde W}=\sum_{m=2}^N\lambda_m|\Psi_m\rangle\langle\Psi_m|$ and ${\tilde 1}= \sum_{m=2}^N |\Psi_m\rangle\langle \Psi_m|$, see also below relation (\ref{rfu}).}

\begin{equation}
 \label{limiting}
 \begin{array}{l}
 \ds D(\lambda) = \lim_{N\rightarrow\infty} \sum_{m=2}^N\delta(\lambda-\lambda_m) \\ \\
 \hspace{0.5cm} \ds = \frac{1}{\pi}\Im\left( tr (\lambda{\tilde 1}-{\tilde W}-i0{\hat 1})^{-1} \right)
 \end{array}
\end{equation}
where $\delta(..)$ denotes Dirac's $\delta$-function
and $|{\tilde \Psi}(\lambda)\rangle$ indicate (appropriately renormalized) eigenfunctions. In (\ref{limiting})  we further introduced
${\tilde W}= {\cal W}-|\Psi_1\rangle\langle\Psi_1|$ and ${\tilde 1}={\hat 1}-|\Psi_1\rangle\langle\Psi_1|$ .
For the general derivations to follow in this section, however,
it is sufficient to write (\ref{infiniteGreen}) for the sake of simplicity as
an infinite sum. Before we consider the FRW let us carefully consider
how properties change when passing from finite to infinite networks.

For finite and infinite networks the geometrical series in (\ref{pro}) and (\ref{infiniteGreen}) {\it always converge for $|\xi|<1$}.
However, on {\it finite networks} the series in (\ref{pro}) is
convergent {\it only for $|\xi|<1$} but always {\it divergent} for $\xi=1$ due to the existence of the largest eigenvalue $\lambda_1=1$ of the transition matrix ${\cal W}$.
It follows then from (\ref{firstpassagepro}) that in a finite network the probability of ever return to the departure node is
$F(0,0,\xi=1)= 1-\frac{1}{P(0,0,\xi=1)} = 1-0$ with divergent $P(0,0,\xi=1)$. It follows recurrence of the walk on finite networks.
To prove recurrence of the random walk the divergence
of $P(\xi\rightarrow 1)$ is a sufficient criteria (due to the presence of eigenvalue $\lambda_1=1$ corresponding to the non-zero stationary distribution $\frac{1}{N}$ on finite networks).
Random walks on finite networks hence are always recurrent \cite{hudges}.
Further we observe in view of (\ref{pro}) with (\ref{diagonalelement}) that in finite connected networks for all nodes $p$ independent
of the departure node $F(p,q,\xi\rightarrow 1) = 1$ (due to the existence of $\lambda_1=1$).
In finite networks any node $p$ (including the departure node) for sure is ever visited.
As a consequence a search strategy based on the random walk in {\it finite} connected networks is always successful.

Depending on the properties of the network this property may change in the case of infinite networks which we will analyze more closely in what follows.
Before we do so let us further analyze above introduced $F$-matrix (\ref{firstpassagepro}) (generating matrix of the first passage probabilities) taking the representation
\begin{equation}
\label{Fmat}
F(p,q,\xi) =  \sum_{n=0}^{\infty}F_n(p,q)\xi^n = \frac{N^{-1}+(1-\xi)(r_{pq}(\xi)-\delta_ {pq})}{N^{-1} +(1-\xi)r_{pp}(\xi)}
\end{equation}
for a finite network of $N$ nodes
where we have introduced
\begin{equation}
 \label{rfu}
\begin{array}{l}
r(\xi)  = \sum_{n=0}^{\infty}\xi^n{\tilde W}^n = [{\tilde 1}-\xi{\tilde W}]^{-1}  , \hspace{1cm} {\tilde W}= {\cal W}-|\Psi_1\rangle\langle\Psi_1| \\ \\
  r_{pq}(\xi)=P(p,q,\xi)-\frac{1}{(1-\xi)}\langle p|\Psi_1\rangle\langle\Psi_1 |q\rangle = \sum_{m=2}^{N}\langle p|\Psi_m\rangle\langle\Psi_m|q\rangle\frac{1}{(1-\lambda_m\xi)}
 \end{array}
\end{equation}
with ${\tilde 1}={\hat 1}-|\Psi_1\rangle\langle\Psi_1|$ indicating the unity in the $N-1$-dimensional subspace of relaxing modes.
Then we have
\begin{equation}
 \label{tracerfu}
 r_{pp}(\xi)= \frac{1}{N}\sum_{q=0}^{N-1}r_{qq}(\xi) = \frac{1}{N}\sum_{m=2}^{N}\frac{1}{(1-\lambda_m\xi)} .
\end{equation}

Since
$r_{pq}(\xi=0)= \delta_{pq}-\frac{1}{N}$ we have $F(p,q,\xi=0)=0$, i.e.  all matrix elements of the zero order in $\xi$ are vanishing due to the fact
that first passage probabilities at $t=0$ are vanishing for all nodes.
Thus series (\ref{Fmat}) $F(p,q,\xi)= \sum_{n=1}^{\infty}F_n(p,q)\xi^n$
starts with the first order in $\xi$ and with $F_1(p,q)= \frac{d}{d\xi}F(p,q,\xi)|_{\xi=0}=P_1(p,q)={\cal W}_{pq}$ recovers the transition matrix
as occupation probabilities coinciding with first passage probabilities at $t=1$.
For infinite networks $N\rightarrow \infty$ relation (\ref{Fmat}) takes the form
\begin{equation}
\label{infinityfirst}
F_{\infty}(p,q,\xi) = \frac{(r_{pq}(\xi)-\delta_ {pq})}{r_{pp}(\xi)}
\end{equation}
with
\begin{equation}
\label{rfu1}
r_{pq}(\xi) = P_{N\rightarrow \infty}(p,q,\xi)= \sum_{m=2}^{\infty}\langle p|\Psi_m\rangle\langle\Psi_m|q\rangle\frac{1}{(1-\lambda_m\xi)}
\end{equation}
which has to be read in the sense of asymptotic integral (\ref{infiniteGreen}).
It follows that the matrices of first passage and occupation probability generating functions $F_{N\rightarrow\infty}(\xi), P_{N\rightarrow\infty}(\xi)$ are fully
determined by the spectral properties of the infinite network Laplacian matrix.
We notice that the matrix $r(\xi)$ of (\ref{rfu1}) at $\xi=1$ is also referred to as the fundamental matrix of the walk \cite{zhang,michelitsch-riascos2017}.

On infinite networks a walk is recurrent only if $r_{pp}(\xi\rightarrow 1)$ is diverging. Otherwise
the infinite spectral sum  $r(\xi=1)$ of (\ref{rfu1}) is converging with
$F_{\infty}(p,p,1)= 1-\frac{1}{r_{pp}(1)} < 1$ where the escape probability which is constant for all nodes in a regular network
$1-F_{\infty}(p,p,1)= \frac{1}{r_{pp}(1)} >0$ is non-zero. Such a walk is transient.

We will analyze in the next section the question of recurrence for the FRW on infinite $d$-dimensional simple cubic lattices
to establish a generalization of Polya's recurrence theorem holding for the entire class of random walks with the same asymptotic power law behavior as the FRW.

Another important characteristics is the mean first passage time (MFPT) indicating the average number of time steps $\langle T_{pq} \rangle$ that a random walker needs starting at $q$ to reach node $p$.
The MFPT with (\ref{Fmat}) is obtained as

\begin{equation}
\label{MFPT}
\begin{array}{l}
 \ds \langle T_{pq} \rangle   = \sum_{n=1}^{\infty}n F_n(p,q)= \frac{d}{d\xi} F(0,0,\xi)|_{\xi=1} \\ \\
 \hspace{0.5cm} \ds  = \lim_{\xi\rightarrow 1-0} \frac{N^{-1}(\delta_{pq}-r_{pq}(\xi)+r_{pp}(\xi))+N^{-1}(1-\xi)(r_{pq}'-r_{pp}')+(1-\xi)^2a(\xi)}{(N^{-1}+(1-\xi)r_{pp}(\xi))^2}
 \end{array}
\end{equation}
where $a(\xi)=r_{pp}(\xi)r_{pq}'(\xi)-(r_{pq}(\xi)-\delta_{pq})r_{pp}'(\xi)$ and $(..)'=\frac{d}{d\xi}(..)$.
For finite networks (\ref{MFPT}) becomes\footnote{Since on finite networks
$r_{pq}(\xi=1) , r_{pq}'(\xi=1), N^{-1}$ are finite.}
\begin{equation}
 \label{mfrtime}
 \begin{array}{l}
 \ds  \langle T_{pq} \rangle   = \sum_{n=1}^{\infty}n F_n(p,q)= \frac{d}{d\xi} F(0,0,\xi)|_{\xi=1} =  N(\delta_{pq}-r_{pq}(1)+r_{pp}(1)) \\ \\
  \hspace{1cm} \ds = N\left(\delta_{pq} +\sum_{m=2}^N\frac{\langle p|\Psi_m\rangle\langle\Psi_m|p\rangle- \langle p|\Psi_m\rangle\langle\Psi_m|q\rangle}{(1-\lambda_m)}\right)
  \end{array}
\end{equation}
which was also obtained earlier \cite{riascos12}\footnote{[where $r_{pq}(\xi)|_{\xi=1}= r_{pq}(e^{-s})|_{s=0}=R^{(0)}_{pq}$ and $R^{(0)}_{pq}$ and $\frac{1}{N}=P_i^{\infty}$
is the notation used in \cite{riascos12}, see eq. (10)]}.
For $p=q$ (\ref{mfrtime}) yields the average number of steps for first return as $\langle T_{pp} \rangle =N$ being constant for all nodes on
a regular network increasing linearly with the number $N$ of nodes and independent of the spectral properties of the Laplacian matrix.
We emphasize that (\ref{mfrtime}) holds for finite connected regular networks ($N<\infty $) with constant degree of the nodes.
For infinite networks (\ref{MFPT}) takes the asymptotic form

\begin{equation}
\label{MFPTinfty}
\ds  \langle T_{pq} \rangle_{\infty} = \lim_{\xi\rightarrow 1}\frac{a(\xi)}{r_{pp}(\xi)}
\end{equation}
where $a(\xi)=r_{pp}(\xi)r_{pq}'(\xi)-(r_{pq}(\xi)-\delta_{pq})r_{pp}'(\xi)$ and $r_{pq}(\xi)=P_{N\rightarrow\infty}(p,q,\xi)$.

A further interesting quantity is the global mean first passage time which is defined as the average value of (\ref{mfrtime}) averaged over all nodes of the network
\begin{equation}
\label{meanglobal}
\langle T \rangle = \frac{1}{N}\sum_{p=0}^{N-1}\langle T_{pq} \rangle  = 1+Nr_{pp}(1) =  1+ \sum_{m=2}^N\frac{1}{(1-\lambda_m)} = 1+K\sum_{m=2}^N\mu_m^{-1}
\end{equation}
where in the last relation we have used
 $\sum_{p=0}^{N-1}r_{pq} = 0$. The global MFPT $\langle T \rangle$ can be interpreted as the average number of time steps to reach any node of the network when starting at a node $q$. We observe
 the remarkable property that (\ref{meanglobal}) does not depend on $q$, for a further general discussion see also \cite{doyle,zhang}.

 When we exclude in the average (\ref{meanglobal}) the contribution of return walks $\langle T_{pp} \rangle$ then we arrive at a global mean first passage time which
 indicates the average number of steps to reach a randomly chosen destination node (different from the departure node) for the first time.
 Without counting the contribution of recurrent walks the global MFPT yields
 \begin{equation}
  \label{kemenycosnat}
  {\cal K}_e =\langle T \rangle -1 =tr({\tilde W})=\sum_{m=2}^N\frac{1}{(1-\lambda_m)} =Nr_{pp}(\xi=1)
 \end{equation}
and is referred to as {\it Kemeny constant} \cite{riascos12,zhang,doyle,kemeny}.
In the picture of diffusive transport phenomena described by the random walk, the inverse Kemeny constant (inverse global MFPT) ${\cal K}_e^{-1}$
measures the speed
of the random walk.
The smaller the Kemeny constant ${\cal K}_e$ the faster the random walker moves threw the network.

\subsection{Some general features and useful formulas for the Fractional Random Walk}

Before analyzing lattices let us briefly deduce in this subsection some useful formulas to analyze the FRW.
The good properties of the fractional Laplacian matrix $L^{\frac{\alpha}{2}}$ are maintained in the interval $0<\alpha\leq 2$\footnote{A
discussion and detailed proof can be found in \cite{michelitsch-riascos2017}.}.
These properties are the following: The zero eigenvalue $\mu_1=0$ and the remaining $N-1$ positive eigenvalues $\mu_m^{\frac{\alpha}{2}} >0$ of the fractional Laplacian matrix
are maintained. The off-diagonal elements $\langle p |L^{\frac{\alpha}{2}}|q\rangle \leq 0 $ remain non-positive as in (\ref{Laplacianmatrix}).
These properties guarantee the existence of the representation $\langle p | L^{\frac{\alpha}{2}} q\rangle =K^{(\alpha)}\delta_{pq}-A^{(\alpha)}_{pq}$ with the positive fractional degree $K^{(\alpha)}$
and non-negative elements of the fractional adjacency matrix (\ref{fractionaldjacncy}).
The fundamental matrix (`Green's function') (\ref{rfu}) at $\xi=1$ takes the form
\begin{equation}
 \label{rfrac}
 r_{pq}^{(\alpha)}(\xi=1) =  K^{(\alpha)}\sum_{m=2}^N|\Psi_m\rangle\langle\Psi_m|(\mu_m)^{-\frac{\alpha}{2}}
\end{equation}
where $K^{(\alpha)}=\frac{1}{N}\sum_{m=2}^N (\mu_m)^{\frac{\alpha}{2}}$ indicates the fractional degree introduced in (\ref{fractionaldegree}).
An interesting representation for the fractional fundamental matrix (\ref{rfrac}) useful in the analysis of FRWs on lattices is obtained in terms of Mellin transforms.
Let $f(\tau)$ an arbitrary bounded function $|f(\tau)| <\infty$ with existing Mellin transform\footnote{Defined on $0\leq \tau<\infty$ and
 decaying  as $\tau\rightarrow \infty$ at least as $f(\tau) \leq const \, \tau^{-\beta}$ with $\beta >\frac{\alpha}{2}$ as $\tau\rightarrow\infty$} which is defined by \cite{abramo}
\begin{equation}
 \label{Mellin}
 {\cal M}_f\left(\frac{\alpha}{2}\right) = \int_0^{\infty} f(\tau) \tau^{\frac{\alpha}{2} - 1}{\rm d}\tau < \infty
\end{equation}
where we always confine us on the admissible range of the fractional Laplacian matrix $0<\alpha\leq 2$.
The fractional fundamental matrix (\ref{rfrac}) can then be represented by Mellin transform of the $N\times N$- matrix function $f(L\tau)$ as
\begin{equation}
 \label{fractfu}
 \frac{r^{(\alpha)}(1)}{K^{(\alpha)}} = \frac{1}{{\cal M}_f(\frac{\alpha}{2})} \int_0^{\infty} \left(f(L\tau)-f(0)|\Psi_1\rangle\langle\Psi_1|\right) \tau^{\frac{\alpha}{2} - 1}{\rm d}\tau
\end{equation}
which is well defined by scalar integrals by employing the spectral representation of the matrix function
$f(L\tau) = f(0)|\Psi_1\rangle\langle\Psi_1| + \sum_{m=2}^Nf(\mu_m\tau) |\Psi_m\rangle\langle\Psi_m|$.
The large choice of functions $f$
opens a convenient tool to generate useful integral representations for the fundamental matrix of the FRW.
By choosing $f(\tau)= e^{-\tau}$ which refers to this class of functions $f$ and  ${\cal M}_{\exp(..)} = \int_0^{\infty}e^{-\tau} \tau^{\frac{\alpha}{2}-1}{\rm d}\tau =\Gamma(\frac{\alpha}{2})$
we get

\begin{equation}
 \label{useful}
  r^{(\alpha)}(1) = \frac{K^{(\alpha)}}{\Gamma(\frac{\alpha}{2})} \int_0^{\infty} \left(e^{-\tau L}\ -|\Psi_1\rangle\langle\Psi_1|\right) \tau^{\frac{\alpha}{2} - 1}{\rm d}\tau
\end{equation}
where $\Gamma(..)$ denotes the $\Gamma$-function
and $e^{-\tau L}=|\Psi_1\rangle\langle\Psi_1|+\sum_{m=2}^N|\Psi_m\rangle\langle\Psi_m|e^{-\mu_m\tau}$ the matrix exponential of Laplacian matrix $L$.
Relation (\ref{useful}) is especially useful as
it links the fundamental matrix of the time-discrete FRW with the matrix exponential $e^{-L t}$ which can be interpreted as the transition matrix of a time-continuous
NRW having a probability distribution evolving as
$\vec{\tilde P}(t)= e^{-Lt} \vec{\tilde P}(0)$ with the master equation $\frac{d}{d t}\vec{\tilde P}(t)= -L \vec{\tilde P}(t)$ \cite{michelitsch-riascos2017}.

Consider now the important limit $\alpha\rightarrow 0+$.
In this limiting case the $N-1$ non-zero eigenvalues take asymptotically $(\mu_m)^{\frac{\alpha}{2}}\rightarrow 1$ and hence we get
for the fractional Laplacian matrix the limiting expression
\begin{equation}
\label{limitingexpr}
\begin{array}{l}
\displaystyle \lim_{\alpha\rightarrow 0+} L^{\frac{\alpha}{2}} = \sum_{m=2}^N |\Psi_m\rangle\langle\Psi_m| = {\hat 1}-|\Psi_1\rangle\langle\Psi_1| ,\\ \\
 \displaystyle \hspace{0.5cm} \lim_{\alpha\rightarrow 0+}\langle p |L^{\frac{\alpha}{2}}|q\rangle = \delta_{pq}\frac{N-1}{N} -\frac{1}{N}(1-\delta_{pq}) .
 \end{array}
\end{equation}
This limiting expression for the fractional Laplacian matrix is universal for finite ergodic regular networks in the sense as it only requires one vanishing and $N-1$ positive eigenvalues $\mu_m$
independent of their values.
Especially we obtain for the limit of the fractional adjacency matrix $A^{(\alpha\rightarrow 0)}_{pq} = \frac{1}{N}(1-\delta_{pq})$
and for fractional degree $K^{(\alpha\rightarrow 0)} = \frac{N-1}{N}$. This leads for $\alpha\rightarrow 0$ to the transition matrix
\begin{equation}
\label{fully}
\lim_{\alpha\rightarrow 0+}{\cal W}^{(\alpha)} =\frac{1}{K^{(\alpha\rightarrow 0)}} A^{(\alpha\rightarrow 0)}_{pq} = \frac{1}{N-1}(1-\delta_{pq})
\end{equation}
coinciding with the transition matrix of a Polya walk on a fully connected network where the walker can reach in one time step any node with equal probability $\frac{1}{N-1}$.
This observation can be confirmed when we return to above 'trivial' scaling invariance of transition matrices of the form (\ref{changesite}). Namely that any rescaled
Laplacian matrix generates an identical random walk. Choosing a scaling factor $\zeta=N$ gives a rescaled Laplacian corresponding to the same random walk as the FRW for $\alpha\rightarrow 0+$ (\ref{fully})
generated by the Laplacian matrix
$N\lim_{\alpha\rightarrow 0}[L^{\frac{\alpha}{2}}]_{pq}=\delta_{pq}(N-1)-(1-\delta_{pq})$ of a {\it fully connected regular network} where the degree of any node is
$K=N-1$. In other words in the limit of vanishing $\alpha\rightarrow 0+$ the FRW exhibits the extreme fast small world dynamics of a Polya type NRW performed on a {\it fully connected network}.

This universal extreme small world property of the FRW for $\alpha\rightarrow 0$ on finite regular networks is independent of
the spectral properties of the Laplacian. This observation underlines the main effect of the FRW dynamics becoming especially pronounced at small $\alpha$ where any network
appears as a small world network and any node can be reached in one time step with equal probability $\frac{1}{(N-1)}$ as in a completely connected network.

In the subsequent section we consider FRWs on lattices with a more profound analysis of some features of the FRW demonstrating the remarkable richness of the FRW.
A key role in this dynamics plays the
fractional scaling index $\alpha$ appearing as a controlling parameter which switches between large world (for $\alpha=2$ corresponding to the NRW) to a small world $0<\alpha<2$ where the small
world becoming extremely pronounced for small $\alpha$.

An important quantity in this analysis is as mentioned the Kemeny constant of the FRW ${\cal  K}_e^{(\alpha)}$
being invariant towards a rescaling with any nonzero scaling factor of the fractional Laplacian. Using relation (\ref{kemenycosnat}) the Kemeny constant (global MFPT) of the FRW can be represented
by the eigenvalues of the
fractional Laplacian as

\begin{equation}
 \label{fractionalkemeny}
 {\cal  K}_e^{(\alpha)} =tr(r^{(\alpha)}(\xi=1)) = \frac{1}{N}\sum_{m=2}^N(\mu_m)^{\frac{\alpha}{2}}\sum_{m=2}^N(\mu_m)^{-\frac{\alpha}{2}}
\end{equation}
taking in the limit for $\alpha\rightarrow 0+$ the value ${\cal  K}_e^{(\alpha\rightarrow 0)} = (N-1)K^{(\alpha\rightarrow 0 )}=\frac{(N-1)^2}{N}$, coinciding with the limiting value obtained
for long-range L\'evy walks when assuming regular networks \cite{riascos12}.

Further let us revisit the inverse of the escape probability in the infinite network limit being defined as the diagonal element $r_{pp}^{(\alpha)}(\xi=1)=\lim_{N\rightarrow\infty}\frac{1}{N}{\cal  K}_e^{(\alpha)}$
being obtained by the limit where $\alpha\rightarrow 0+$

\begin{equation}
\label{extreme}
r_{pp}^{(\alpha)}(\xi=1)=\lim_{N\rightarrow\infty}
 \frac{1}{N}\sum_{m=2}^{N}(\mu_m)^{\frac{\alpha}{2}}\frac{1}{N}\sum_{m=2}^N(\mu_m)^{-\frac{\alpha}{2}} = \lim_{N\rightarrow\infty} \frac{(N-1)^2}{N^2} =1
\end{equation}
i.e. for infinite networks for $\alpha\rightarrow 0+$ $(r_{pp}^{(\alpha \rightarrow 0+)}(\xi=1))^{-1}=1$, i.e. the walker is sure to escape. This property of extreme transience at $\alpha=0+$ of the FRW
is independent on the spectral details
of the Laplacian matrix and therefore universal.
It remains true for a FRW on a infinite $d$-dimensional lattice for any lattice dimension $d$.
For any infinite network the limit $\alpha\rightarrow 0+$ represents a limit of
extreme transience.

This picture is consistent with above mentioned interpretation of the components of the Green's function  $r_{pq}^{(\alpha)}(\xi=1)$ (see below eq. (\ref{probabgen})), namely as the mean residence times (MRT) of the walker in a node: In the limit of extreme transience $\alpha\rightarrow 0+$
with $r_{pp}^{(\alpha\rightarrow 0+)}=1$ indicates that the walker in the average is present in the departure node only once, namely at the time of its departure $t=0$, and never returning to the departure node.
We will return to the important property of extreme transience in the subsequent section when analyzing recurrence of FRWs on $d$-dimensional infinite lattices.

\section{Fractional random walks on simple $d$-dimensional cubic lattices}
\label{sec2}

We now consider
periodic $d$ dimensional periodic lattices ($d$-tori) where $d=1,2,3,4,..$ can take any integer dimension. We assume that the lattice points represent the nodes of the network
and denote them by
$\vec{p}=(p_1,..,p_d)$. In each dimension $j=1,..,d$ we denote the nodes by
$p_j=0,..,N_j-1$ where the total number of nodes of the network is $N=\prod_{j=1}^d N_j$. We assume that the lattice is $N_j$-periodic in each spatial dimension $j=1,..,d$.
All quantities defined on the nodes such as occupation- and first passage probabilities fulfill the periodicity conditions $Q_(p_1,..,p_j,..,p_n)= Q_(p_1,..,p_j+N_j,..,p_n)$.

It follows that the matrices defined above all have the same canonic basis of eigenvectors
$|\Psi_{\vec{\ell}}\rangle=\prod_{j=1}^d|\Psi_{\ell_j}\rangle$ where $\vec{\ell}=(\ell_1,..,\ell_d)$ $\ell_j=0,..,N_j-1$
 is represented by the Bloch eigenvectors with the notation $\langle\vec{p}|\Psi_{\vec{\ell}}\rangle =
\prod_{j=1}^d \frac{e^{i\kappa_{\ell_j}p_j}}{\sqrt{N_j}} = \frac{e^{i{\vec{p}\cdot\vec{\kappa}_{\vec{\ell}}}}}{\sqrt{N}} $.
In order to fulfill $N_j$-periodicity it follows that the components of wave vectors $\vec{\kappa}=(\kappa_{\ell_j})$ of the Bloch eigenvectors can can only take the values
$\kappa_{\ell_j} =\frac{2\pi}{N_j}\ell_j$, $\ell_j=0,..,N_j-1$ ($j=1,..,d$).
\newline\
We plotted $d$-dimensional periodic lattices for dimensions $d=1,2$ in Figure 1.
Topologically such a lattice can be conceived
as $d$-dimensional hypertorus (`$d$-torus'), for instance in 1D this is a cyclic ring, in 2D a conventional torus, and so forth.
\begin{figure*}[!h]
\begin{center}
\includegraphics*[width=0.76\textwidth]{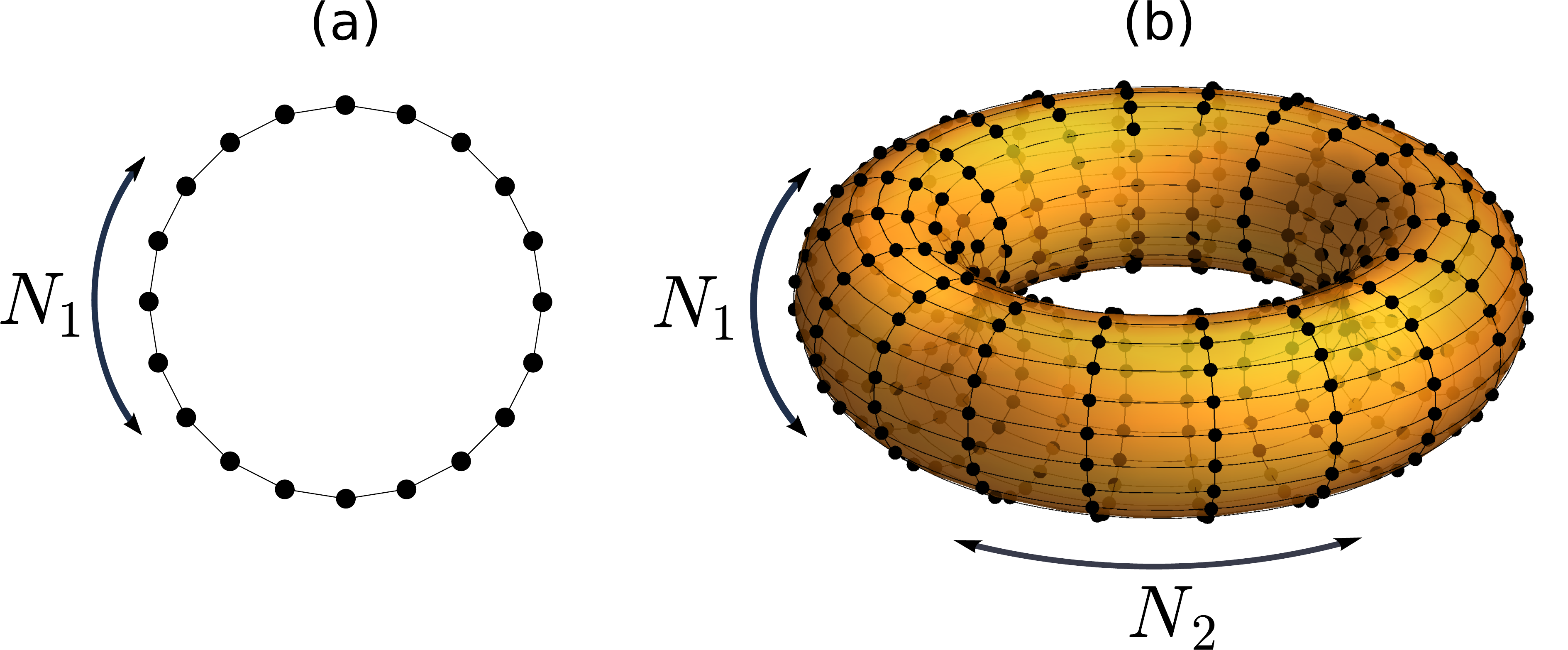}
\end{center}
\vspace{-8mm}
\caption{\label{Figure1} Finite lattices with periodic boundary conditions. (a) One-dimensional lattice with length $N_1=20$,
the resulting structure is a ring. (b) Two-dimensional lattice with dimensions $N_1=N_2=20$,
in this case the nodes define a torus obtained from the Cartesian product of two rings with dimensions $N_1$, $N_2$. }
\end{figure*}
\newline
\\[2mm]
The spectral decomposition
$B_{pq}= b_1\langle p|\Psi_1\rangle\langle\Psi_1|q\rangle + \sum_{m=2}^N b_m \langle p|\Psi_m\rangle\langle\Psi_m|q\rangle $ of a matrix $B$ on the lattice
has the representation
\begin{equation}
 \label{matwriteas}
 B_{\vec{p}-\vec{q}} = B(p_1-q_1,..,p_d-q_d) =  \sum_{\vec{\ell}} b_{\vec{\ell}}\,  \frac{e^{i{(\vec{p}-\vec{q})\cdot\vec{\kappa}_{\vec{\ell}}}}}{N}  =
 \sum_{\ell_1=0}^{N_1-1}..,\sum_{\ell_j=0}^{N_j-1}\sum_{\ell_d=0}^{N_d-1} b_{(\ell_1,..,\ell_d)}\prod_{j=1}^d \frac{e^{i\kappa_{\ell_j}(p_j-q_j)}}{N_j}
\end{equation}
[where we use alternatively the notations $B_{\vec{p}-\vec{q}}=B(\vec{p}-\vec{q})$].
We identify the vanishing eigenvalue $\mu_1=\mu_{\vec{0}}=0$ of the stationary eigenvector $\langle \vec p |\Psi_{\vec 0}\rangle = \frac{1}{\sqrt{N}} $ corresponding to
the zero (Bloch-wave) vector $\vec{\kappa}_{\ell} = (0,..,0)$. We see from this relation that
such matrices are of the form $B_{\vec{p}|\vec{q}} = B(\vec{p}-\vec{q})$ reflecting translational invariance, and all $N\times N$-matrices fulfill $N_j$-periodic boundary conditions
$B(\vec{s})=B(s_1,..,s_j,..,s_d)=B(s_1,..,s_j+N_j,..,s_d)$
in all dimensions $j=1,..,d$
and further symmetries such as (generalized) T\"oplitz structure have been outlined elsewhere
\cite{michelitsch-riascos2017,michelJphysA,Zoia2007}. Further useful is the transition to infinite lattices when all $N_j\rightarrow\infty$ which we write compactly
\footnote{$\kappa_{\ell_j} =\frac{2\pi}{N_j}\ell_j \rightarrow \kappa_j$ thus the eigenvalues become continuous functions
$ b_{(\ell_1,..,\ell_d)} =b(\kappa_1,..,\kappa_d)=b(\vec{\kappa})$ where $0\leq \kappa_j\leq 2\pi$ and $\frac{1}{N_j}=\frac{{\rm d}\kappa_j}{2\pi}$.}
\begin{equation}
 \label{infinitespectral}
 B(\vec{p}) =\frac{1}{(2\pi)^d} \int b(\vec{\kappa}) e^{i{\vec{p}\cdot\vec{\kappa}_{\vec{\ell}}}} {\rm d}^d\kappa =:
 \frac{1}{(2\pi)^d} \int_{-\pi}^{\pi}{\rm d}\kappa_1..\int_{-\pi}^{\pi}{\rm d}\kappa_d \, b(\kappa_1,..,\kappa_d) e^{i{\vec{p}\cdot\vec{\kappa}_{\vec{\ell}}}} .
\end{equation}
Especially we identify in the context of lattices the general representation of the components of the unity matrix
$\delta_{pq}\rightarrow \delta_{\vec{p}-\vec{q}} = \prod_{j=1}^d\delta_{p_jq_j}$.

Let us now introduce the fractional Laplacian matrix on the $d$-dimensional lattice  which we generate from a $N\times N$-Laplacian matrix ${\cal L}$ defined on the lattice
with next neighbor connections \cite{michelitsch-riascos2017}
\begin{equation}
 \label{explicit}
{\cal L}(\vec{p}-\vec{q})= {\cal L}_{p_1,..p_n|q_1,..,q_n} =2d \prod_{j=1}^d\delta_{p_jq_j} - \sum_{j=1}^d\left(\delta_{p_{j+1}q_j}+ \delta_{p_{j-1}q_j} \right)\prod_{s\neq j}^n \delta_{p_sq_s}
\end{equation}
where the constant degree of the $d$-dimensional lattice is the number of next neighbor nodes $K=2d$.
The spectral representation of the fractional Laplacian on the finite lattice
 ${\cal L}^{\frac{\alpha}{2}}$ is
\begin{equation}
\label{spectralfracb}
[{\cal L}^{\frac{\alpha}{2}}]_{\vec{p}\vec{q}} = [{\cal L}^{\frac{\alpha}{2}}]({\vec{p}-\vec{q}})
=\frac{1}{N}
\sum_{\vec{\ell}} e^{i\vec{\kappa}_{\vec{\ell}}\cdot(\vec{p}-\vec{q})}\mu_{\vec{\ell}}^{\frac{\alpha}{2}} ,
\hspace{0.5cm} \mu(\vec{\kappa}_{\vec{\ell}}) = 2d-2\sum_{j=1}^d\cos{(\kappa_{\ell_j})} ,\hspace{0.5cm} 0< \alpha \leq 2
\end{equation}
where throughout this analysis we confine us on the admissible range $0<\alpha \leq 2$ where $\alpha=2$ recovers (\ref{explicit}).
The fractional transition matrix on the lattice which we define as in (\ref{FRWtrans})
then writes for the finite lattice
\begin{equation}
 \label{transitionfractional}
 {\cal W}^{(\alpha)}(\vec{p}-\vec{q}) = \sum_{\vec{\ell}} \lambda_{\vec{\ell}}^{(\alpha)} \frac{e^{i{(\vec{p}-\vec{q})\cdot\vec{\kappa}_{\vec{\ell}}}}}{N}
 ,\hspace{0.5cm} \lambda_{\vec{\ell}}^{(\alpha)} = 1- \frac{\mu^{\frac{\alpha}{2}}(\vec{\kappa}_{\vec{\ell}})}{K^{(\alpha)}}
\end{equation}
where $K^{(\alpha)}=\frac{1}{N}tr({\cal L}^{\frac{\alpha}{2}})= \frac{1}{N}\sum_{\vec{\ell}}\mu_{\vec{\ell}}^{\frac{\alpha}{2}}$ indicates the fractional degree. We notice further
that the diagonal elements of the fractional transition matrix are vanishing due to
$\sum_{\vec{\ell}} \lambda_{\vec{\ell}}^{(\alpha)} = 0$ forcing the fractional random walker to change the node at each step.

We analyze now infinite lattices $N_j\rightarrow\infty$ ($\forall j=1,..,d$) to consider the question of recurrence of the FRW where we employ the limiting formula
(\ref{infinitespectral}) for the spectral representations. Let us first analyze
the probability $F^{(\alpha)}(p,q,\xi=1)$
of ever passage which is determined by\footnote{where all above general subscripts are adopted to $(..)_{pq} \rightarrow \vec{p}-\vec{q}$.}

\begin{equation}
 \label{fractionaleverret}
  F^{(\alpha)}_{\vec{p}-\vec{q}}(\xi=1) = \frac{r^{(\alpha)}_{\vec{p}-\vec{q}}(1)-\delta_{\vec{p}-\vec{q}}}{r^{(\alpha)}_{\vec{0}}(1)} .
\end{equation}
In order to analyze recurrence it is sufficient to consider the diagonal element of (\ref{fractionaleverret}) which indicates
the probability that the walker {\it ever} returns to the departure node
\begin{equation}
 \label{everreturn}
 F^{(\alpha)}_{\vec{0}}(\xi=1) = 1-\frac{1}{r^{(\alpha)}_{\vec{0}}(\xi=1)} \leq 1 .
\end{equation}
It follows that $1-\ F^{(\alpha)}_{\vec{0}}(\xi=1)=\frac{1}{r^{(\alpha)}_{\vec{0}}(\xi=1)}$
indicates the (escape-) probability that the walker {\it never} returns to its departure node.
In (\ref{fractionaleverret}), (\ref{everreturn}) we may use general relation (\ref{rfu1}) which writes for the infinite $d$-dimensional lattice
\begin{equation}
 \label{rpqalpha}
 r^{(\alpha)}_{\vec{p}-\vec{q}}(\xi=1) = \frac{K^{(\alpha)}}{(2\pi)^d}\int e^{i{(\vec{p}-\vec{q})\cdot\vec{\kappa}_{\vec{\ell}}}} \mu^{-\frac{\alpha}{2}}(\vec{\kappa}) {\rm d}^d\kappa
\end{equation}
which contains the fractional degree
\begin{equation}
 \label{fractionaldegree2}
 K^{(\alpha)} = \frac{1}{(2\pi)^d} \int \mu^{\frac{\alpha}{2}}(\vec{\kappa}) {\rm d}^d\kappa ,\hspace{1cm}\mu(\vec{\kappa})= \left(2d-2\sum_{j=1}^d\cos{(\kappa_j)}\right)
\end{equation}
and where the eigenvalues of the fractional Laplacian $\mu^{\frac{\alpha}{2}}(\vec{\kappa})$
become a continuous function defined on the $d$-dimensional cube (Brillouin zone) of volume $(2\pi)^d$.

We analyze now recurrence (transience) of the FRW.
To this end it is sufficient to consider the existence of identical diagonal elements of (\ref{rpqalpha})

\begin{equation}
 \label{relation}
 r^{(\alpha)}_{\vec{0}}(\xi=1) = \frac{1}{(2\pi)^{2d}} \int  \mu^{\frac{\alpha}{2}}(\vec{\kappa}) {\rm d}^d\kappa
 \int  \mu^{-\frac{\alpha}{2}}(\vec{\kappa'}) {\rm d}^d\kappa' .
\end{equation}

We notice that (\ref{relation}) is related with the Kemeny constant (\ref{fractionalkemeny})
(global MFPT) by $ r^{(\alpha)}_{\vec{0}}(\xi=1) = \lim_{N\rightarrow\infty} \frac{{\cal K}_e^{(\alpha)}}{N}$
and hence the probability of never return for the walker performing the FRW $ (r^{(\alpha)}_{\vec{0}}(\xi=1))^{-1} \sim \frac{N}{{\cal K}_e^{(\alpha)}}$ is the infinite network limit of the
inverse of the global MFPT.

In above considerations we have seen that $\frac{1}{r^{(\alpha)}_{\vec{0}}(\xi=1)}$ indicates the escape probability, i.e. probability of never return of the walker
to the departure node. As mentioned the FRW is hence {\it recurrent} only if $r^{(\alpha)}_{\vec{0}}(\xi=1)\rightarrow \infty$ is divergent and {\it transient} otherwise.
Since the integral (\ref{fractionaldegree2}) for the fractional degree in the admissible $\alpha$ range always exists,
the question of recurrence depends uniquely
on the divergence (or convergence) of the second integral in (\ref{relation}) depending crucially on the features
of the $\mu^{-\frac{\alpha}{2}}$ for small $|\vec{\kappa}| \rightarrow 0 $ arround the origin.

Taking into account that the eigenvalues (\ref{spectralfracb}) of the fractional Laplacian matrix
arround the origin are behaving as $\mu^{\frac{\alpha}{2}}({\vec{\kappa}}) \sim \kappa^{\alpha}$
($\kappa=|\vec{\kappa}|=\sqrt{\sum_{j=1}^d\kappa_j^2}$), then $r^{(\alpha)}_{\vec{0}}(\xi=1)$ of (\ref{relation}) and the matrix elements (\ref{rpqalpha}) are finite if
\footnote{The appearance
of the additional factor $\kappa^{d-1}$ in the integrand is due to scaling of the volume element
${\rm d}^d\vec{\kappa} = \kappa^{d-1}{\rm d}\kappa {\rm d}\Omega_d$ and $\int_{\kappa=1} {\rm d}\Omega_d = \frac{2\pi^{\frac{d}{2}}}{\Gamma(\frac{d}{2})}$ is the surface of the $d$-dimensional unit ball.}

\begin{equation}
 \label{leadingcon}
 \begin{array}{l}
 \displaystyle
 \frac{r^{(\alpha)}_{\vec{0}}(\xi=1)}{K^{(\alpha)}} = \frac{1}{(2\pi)^{d}}  \int \mu^{-\frac{\alpha}{2}}(\vec{\kappa'}) {\rm d}^d\kappa' \\ \\
\displaystyle \hspace{1.5cm} = \frac{1}{(2\pi)^{d}} \frac{2\pi^{\frac{d}{2}}}{\Gamma(\frac{d}{2})} \lim_{\epsilon\rightarrow 0} \int_{\epsilon}^{\kappa_0}
 \kappa^{d-1-\alpha}{\rm d}\kappa + \int_{V_c} \mu^{-\frac{\alpha}{2}}(\vec{\kappa'}) {\rm d}^d\kappa'   \sim \lim_{\epsilon\rightarrow 0} a(\epsilon)  + C(\kappa_0)
 \end{array}
\end{equation}
exists. In (\ref{leadingcon}) $0<\kappa_0 << 1$ is sufficiently small that
$(\mu(\vec{\kappa_0}))^{-\frac{\alpha}{2}} \approx \kappa_0^{-\alpha}$ and $C(\kappa_0)$ is the contribution of the integral of $\mu^{-\frac{\alpha}{2}}(\vec{\kappa'})$ over $V_c$ which is
the cube $-\pi<\kappa_j<\pi$ without the $d$-sphere of radius $\kappa=\kappa_0$.

\begin{figure*}[!t]
\begin{center}
\includegraphics[width=0.4\textwidth]{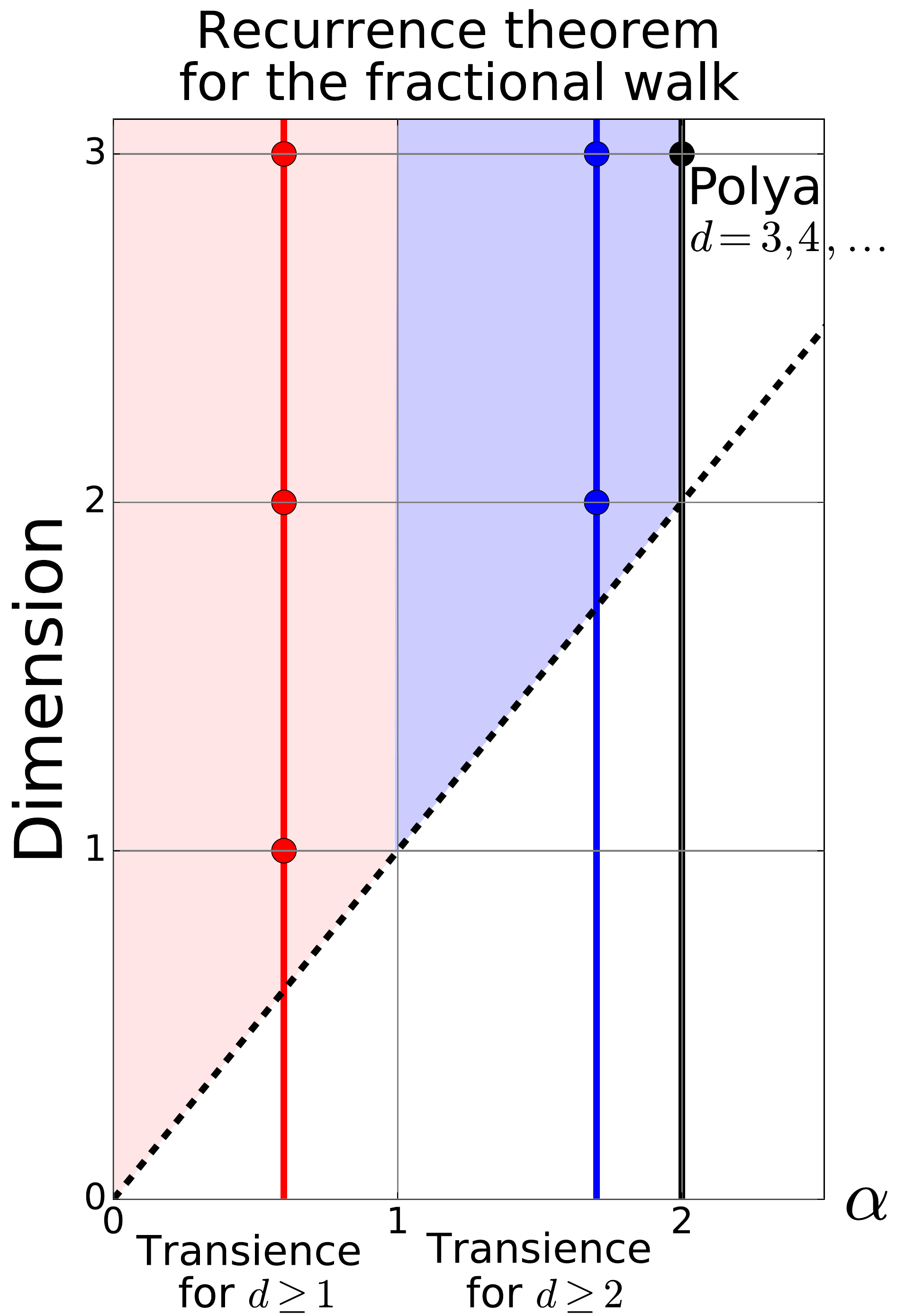}
\end{center}
\vspace{-5mm}
\caption{\label{FigureNew} Representation of the recurrence theorem for the Fractional Random Walk for the admissible range $0<\alpha\leq 2$: The plot shows for two values of $\alpha$ (within $0<\alpha<2$)  the lattice dimensions of transience (indicated as bullet points). Recurrent FRWs exist only within $1\leq \alpha < 2$ for $d=1$.
For $\alpha=2$ representing the Polya walk, Polya's classical recurrence theorem is recovered where the walk is recurrent
for lattice dimensions $d=1,2$ and transient for $d\geq 3$.
}
\end{figure*}

The first integral (\ref{leadingcon}) is crucial for the divergence or convergence of $r^{(\alpha)}_{\vec{0}}$:
It
behaves as
$a(\epsilon)\sim \frac{\epsilon^{d-\alpha}}{d-\alpha} $ for $d\neq\alpha$ and $a(\epsilon) \sim log(\epsilon)$ when $d=\alpha$ where $\epsilon\rightarrow 0+$.
Hence (\ref{leadingcon}) diverges for $d\leq \alpha$ and as a consequence the FRW then is recurrent.
On the other hand integral (\ref{leadingcon}) is finite for $d>\alpha$ and as a consequence the FRW then is transient where the walker has a finite escape probability (probability of never return to the departure node) $\frac{1}{ r^{(\alpha)}_{\vec{0}}(\xi=1)}$.
\newline\newline
{\it The generalized recurrence theorem for Fractional Random Walks can hence be formulated as follows. The FRW is recurrent for lattice dimensions $d\leq \alpha$ and transient for $d>\alpha$ where always
$0<\alpha \leq 2$. This behavior is represented in Figure \ref{FigureNew}: Lattice dimensions of transient FRWs are indicated by bullet points.
The recurrence theorem remains true for the entire class of random walks on infinite networks with the same power law asymptotics as the FRW leading to the emergence of L\'evy flights. }
\newline\newline
The recurrence theorem holds also for L\'evy flights in the continuous $d$-dimensional infinite space
\cite{sato} (see chapter 7, pages 261, 262).
For L\'evy flights on lattices, transience has been demonstrated for lattice dimensions $d\geq 2$ ($0<\alpha<2$) in \cite{ferraro}.

Generally it is sufficient to consider the infinite space Green's function
     which yields
     a Riesz potential $\lim_{\epsilon\rightarrow 0+}\frac{1}{(2\pi)^d}\int e^{i{\vec k}\cdot\vec{x}}|\vec{k}|^{-\alpha}e^{-\epsilon|k|}{\rm d}^d\vec{k} = \frac{-C_{-\alpha,d}}{|\vec{x}|^{d-\alpha}}$
     for $d>\alpha$ (see Eq. (\ref{rieszddim})) and is divergent for $d\leq\alpha$.

\vspace{1cm}


\noindent As $0<\alpha\leq 2$ always is restricted only the following cases exist:\newline\newline
\noindent {\bf (i)} $0<\alpha < 1$: $d-\alpha >0 \forall d$ where $(r^{(\alpha)}_{\vec{0}}(\xi\rightarrow 1))^{-1} >0$ is a nonzero escape
probability. In this range the FRW is {\it transient} for all lattice dimensions $d$.
We therefore refer the interval $0<\alpha < 1$ to as `{\it strongly transient regime}' (see Figure \ref{FigureNew}). The transience becomes more and more pronounced the smaller $\alpha$.
This includes the above discussed limiting
case of extreme
transience (\ref{extreme}). \newline\newline
\noindent {\bf (ii)}  $1\leq\alpha < 2 $: $d-\alpha > 0$ $(r^{(\alpha)}_{\vec{0}}(\xi\rightarrow 1))^{-1} >0$,  i.e. we have nonzero escape probability with transience
for $d=2,3,..$, recurrence for $d=1$ (Figure  \ref{FigureNew}).\newline\newline
\noindent {\bf (iii)} $\alpha =2$ (Polya walk): (\ref{leadingcon}) diverges for dimensions $d=1,2$, whereas converges for
dimensions greater than $d=3,4,..$.
This recovers Polya's classical recurrence theorem \cite{polya}: The Polya walk is recurrent for dimensions $d=1,2$ and transient for dimensions $d>2$ (Figure  \ref{FigureNew}).
\newline\newline
The statements (i)-(iii) generalize Polya's recurrence theorem to Fractional Random Walks.
In appendix \ref{appendb} we give a brief demonstration for the emergence of L\'evy flights for FRWs within $0<\alpha<2$
(and Brownian motion for the Polya case $\alpha=2$)
 on
infinite lattices due to the power law asymptotics of the eigenvalues. The same asymptotic behavior is also responsible for the convergence or divergence of (\ref{leadingcon}) determining transience or
recurrence of the FRW.
The recurrence theorem for the Fractional Random Walk hence remains true for the whole class of
random walks with asymptotic emergence of L\'evy flights.
These are walks generated by Laplacian matrices where the eigenvalues behave asymptotically as a power law $ \sim \kappa^{\alpha}$ when $\kappa\rightarrow 0$ with asymptotic
behavior of the transition matrix elements as for $|\vec{p}-\vec{q}| >>1$ as the kernel of the fractional Laplacian operator
$\sim |\vec{p}-\vec{q}|^{-(d+\alpha)}$, leading {\it in the transient regime $d>\alpha$} to the ever passage probabilities (and lattice Green's functions) decaying as Riesz potentials
$ \sim |\vec{p}-\vec{q}|^{-(d-\alpha)}$ (see appendix \ref{appendb} and \cite{ortiguera}).
\newline

A physical interpretation is as follows. (i) In the interval $0<\alpha< 1$ the smaller $\alpha$ the `faster' is the FRW: Due to the slower decay of the transition matrix elements as
$ |\vec{p}-\vec{q}|^{-(d+\alpha)}$ long range jumps are more frequent leading to strong transience in case (i). With increasing $\alpha$ in case (ii), i.e. for $1 \leq \alpha< 2$ the FRW
becomes slower than in case (i) (the stronger decay of the transition matrix elements make long-range jumps more rare to take place). This relative slowness of the FRW can only be compensated
when $d$ increases (thus transience only for dimensions $d=2,3,..$ in case (ii)).
This tendency is even more pronounced in case (iii)
of the Polya walk where no long-range jumps occur thus transience occurs only for dimensions $d=3,4,..$. The slower the FRW (the larger $\alpha$) the higher the minimum dimension $d$ must be that the walk
becomes transient, reflecting the effect that higher dimensions offer to the random walker more escape paths.

In conclusion the transience of a FRW at $d-\alpha >0$ is the more pronounced
by higher spatial dimensions $d$ (more escape paths)
and lower exponents $\alpha$ increasing the speed of the FRW, that is the escape probability (in the infinite lattice limit $N\rightarrow\infty$)
$(r_{\vec{0}}(1)))^{-1} \sim \frac{N}{\cal K^{(\alpha)}}$ increases when the (renormalized) global MFPT (Kemeny constant) decreases. This behavior is especially pronounced in the limiting case of extreme transience when $\alpha\rightarrow 0+$
where the global MFPT becomes infinitely long
$K^{(\alpha\rightarrow 0)}\sim N\rightarrow \infty$ for large $N$  (see relation (\ref{extreme})).
[This can be seen in view of (\ref{mfrtime}) $N=\langle T_{\vec{0}}\rangle$ indicating the average number of steps for first returns
and the quantity $({\cal K}_e^{(\alpha)})^{-1}$ being a measure for the speed of the FRW].
We mention that the recurrence behavior becomes different for L\'evy walks \cite{barkai} where the velocity of the walker is finite \cite{klafter}.

\section{Transient regime $0<\alpha<1$ for the infinite ring}
\label{explcit1Dring}

We saw by above consideration that in case (i), i.e. for $0<\alpha<1$ the FRW is transient for
all dimensions $d$ of the lattice. Let us now analyze for the transient regime the limiting case $N\rightarrow \infty$ of a cyclic ring ($d=1$) which allows explicit evaluations for the
fractional lattice Green's function.
The elements of the fractional Laplacian matrix have been obtained in closed form
\cite{michelJphysA,michelchaos,Zoia2007}. For a cyclic ring the Laplacian matrix (\ref{explicit}) takes the simple representation
\begin{equation}
 \label{laplacianring}
 L_{pq} = 2\delta_{pq}-\delta_{p,q+1}-\delta_{p,q-1}
\end{equation}
of a symmetric second difference operator where the eigenvalues of the Laplacian for the infinite ring are
\begin{equation}
 \label{infiniteeig}
 \mu(\kappa)= 2(1-\cos(\kappa))= 4 \sin^2\left(\frac{\kappa}{2}\right) ,\hspace{1cm} -\pi \leq \kappa\leq \pi .
\end{equation}
The fractional Laplacian matrix $L^{\frac{\alpha}{2}}$ for the infinite ring has the spectral representation where we account for $(L^{\frac{\alpha}{2}})_{p-q} = (L^{\frac{\alpha}{2}})_{|p-q|}$
\begin{equation}
 \label{spectralrepresentlap}
 (L^{\frac{\alpha}{2}})_{|p|}=\frac{1}{2\pi}\int_{-\pi}^{\pi}e^{i\kappa p}\left(4 \sin^2\frac{\kappa}{2})\right)^{\frac{\alpha}{2}}{\rm d}\kappa .
\end{equation}
The matrix elements of (\ref{spectralrepresentlap}) have been obtained in the following representation \cite{michelJphysA,michelchaos} where the
T\"oplitz structure of this matrix allows $|p-q| \rightarrow |p|=p$

\begin{equation}
 \label{spetraleval}
 \begin{array}{l}
 \displaystyle (L^{\frac{\alpha}{2}})_{|p|}=\frac{2^{\alpha}}{\sqrt{\pi}(p-\frac{1}{2})!} \int_0^1\xi^{\frac{\alpha}{2}}\frac{d^p}{d\xi^p}\left(\xi(1-\xi)\right)^{p-\frac{1}{2}}{\rm d}\xi \\ \\
\displaystyle  \hspace{1cm} = (-1)^p\frac{\alpha!}{(\frac{\alpha}{2}+p)!(\frac{\alpha}{2}-p)!} = -\frac{\alpha !}{\pi}\sin{\left(\frac{\alpha\pi}{2}\right)}\frac{(|p|-1-\frac{\alpha}{2})!}{(\frac{\alpha}{2}+|p|)!} ,
 \end{array}
\end{equation}
where
we utilized the notation with generalized binomial coefficients \cite{michelJphysA} and generalized
factorials holding for (non-negative) integers and non-integers $\zeta!=\Gamma(\zeta+1)$ where $\Gamma(..)$ denotes the $\Gamma$-function\footnote{The second expression in (\ref{spetraleval})
is obtained by the first employing
Euler's reflection formula $\Gamma(z)\Gamma(1-z) = \frac{\pi}{\sin{\pi z}}$ and was also reported in \cite{Zoia2007}.}.
Detailed derivations and discussions of the properties of the explicit form of 1D fractional Laplacian is performed in \cite{michelJphysA}. In view of expression (\ref{spetraleval})
let us consider the sign of the matrix elements (\ref{spetraleval}) first for $p\neq 0$ the sign is determined by
$sgn[(-1)^p\frac{\frac{\alpha}{2}!}{(\frac{\alpha}{2}-p)!}]=sgn[(-1)^p\prod_{s=0}^{p-1}(\frac{\alpha}{2}-s)] = (-1)$ whereas $+1$ for $p=0$. Thus we confirm what we mentioned above
that the diagonal element
($p=0$) of the fractional Laplacian (the fractional degree) is positive whereas the off diagonal elements are all negative within the range $0<\alpha<2$ thus
the fractional Laplacian matrix
constitutes a good generating matrix for a random walk.

As outlined in the previous section the integral of the fundamental matrix $r^{(\alpha)}(\xi=1)$ converges in the transient regime $0<\alpha<1$ for all lattice dimensions $d$: We evaluate now
(\ref{rpqalpha}) for $d=1$ in explicit form. $L^{-\frac{\alpha}{2}}$ is obtained by formally replacing $\alpha\rightarrow-\alpha$ in (\ref{spetraleval}), however, we verify this necessary property
by a brief explicit calculation.

For the 1D infinite ring this integral is determined by the inverse fractional Laplacian matrix (Fractional lattice Green's function)
$r^{(\alpha)}(\xi=1) = K^{(\alpha)}L^{-\frac{\alpha}{2}}$ defined in (46). In contrast to finite networks the fractional Laplacian matrix $L^{\frac{\alpha}{2}}$ becomes invertible in the transient regime
in the limit of infinite lattices $N\rightarrow\infty$ since then ${\hat 1}-|\Psi_1\rangle\langle\Psi_1| \rightarrow {\hat 1}$
due to suppression of the stationary distribution $|\Psi_1\rangle\langle\Psi_1|=\frac{1}{N} \rightarrow 0$. For the 1D infinite ring
$r_{pq}^{(\alpha)}(\xi=1)=r_{|p-q|}^{(\alpha)}(\xi=1)$ exists and is obtained as (where we denote always $r_{pq}=r_{|p-q|}$ and $|p-q| \rightarrow p$)

\begin{equation}
\label{fundmat}
\begin{array}{l}
\displaystyle r_{|p|}^{(\alpha)}(\xi=1) = \frac{K^{(\alpha)}}{2\pi}\int_{-\pi}^{\pi}e^{i\kappa p}\left(4 \sin^2\left(\frac{\kappa}{2}
\right)\right)^{-\frac{\alpha}{2}}{\rm d}\kappa \\ \\
\hspace{1cm} = K^{(\alpha)}
\displaystyle \frac{2^{-\alpha}}{\sqrt{\pi}(p-\frac{1}{2})!} \int_0^1\xi^{-\frac{\alpha}{2}}\frac{d^p}{d\xi^p}\left\{\xi(1-\xi)\right\}^{p-\frac{1}{2}}{\rm d}\xi ,\hspace{1cm} 0<\alpha <1
\end{array}
\end{equation}
where we emphasize again that this integral converges only in the transient regime $0<\alpha<1$. In (\ref{fundmat}) $K^{(\alpha)}$ indicates
the fractional degree being the diagonal element of (\ref{spetraleval}) which yields

\begin{equation}
\label{diagelefrac}
K^{(\alpha)} =  (L^{\frac{\alpha}{2}})_{0} = \frac{2^{\alpha}}{\pi}\frac{(\frac{\alpha-1}{2})! (-\frac{1}{2})!}{\frac{\alpha}{2}!} = \frac{\alpha!}{\frac{\alpha}{2}!\frac{\alpha}{2}!} .
\end{equation}

The matrix elements
(\ref{fundmat}) can be evaluated in the same way
as (\ref{spetraleval}): Upon $p=|p|$ partial integrations (\ref{fundmat}) yields
\begin{equation}
\label{fundmat2}
\begin{array}{l}
\displaystyle r_{|p|}^{(\alpha)}(\xi=1) = K^{(\alpha)}\frac{2^{-\alpha}}{\sqrt{\pi}(p-\frac{1}{2})!} \left((-1)^p\prod_{s=0}^{p-1}(-\frac{\alpha}{2}-s)\right)\int_0^1\xi^{-\frac{\alpha+1}{2}}(1-\xi)^{(p-\frac{1}{2})}
{\rm d}\xi \\ \\
\displaystyle \hspace{1cm} =  K^{(\alpha)}\frac{2^{-\alpha}(\frac{(-\alpha-1)}{2})!}{\sqrt{\pi}(-\frac{\alpha}{2}+p)!} (-1)^p\frac{(\frac{-\alpha}{2})!}{(\frac{-\alpha}{2}-p)!}   ,\hspace{1cm} 0<\alpha <1
\end{array}
\end{equation}
where $(-1)^p\prod_{s=0}^{p-1}(-\frac{\alpha}{2}-s)= \prod_{s=0}^{p-1}(\frac{\alpha}{2}+s)= (-1)^p\frac{(\frac{-\alpha}{2})!}{(\frac{-\alpha}{2}-p)!} >0$ and
hence (\ref{fundmat2}) is {\it uniquely positive}. The first relation (\ref{fundmat2})$_1$ is written for $p\neq 0$.
For $p=0$ the product $(-1)^p\prod_{s=0}^{p-1}(..)$ has to be replaced by $1$ whereas the second equation (\ref{fundmat2})$_2$
holds for all components $|p|=0,1,2,..$ including $p=0$ where in all expressions we write $p=|p|$.
Using the identity\footnote{For details, see again \cite{michelJphysA}.}
 $\frac{2^{-\alpha}(\frac{-(\alpha+1)}{2})!}{\sqrt{\pi}(\frac{-\alpha}{2})!}=
 \frac{(-\alpha)!}{(\frac{-\alpha}{2})!(\frac{-\alpha}{2})!}$ finally yields for (\ref{fundmat2})
 the more handy expression
\begin{equation}
\label{fundmat3}
\begin{array}{l}
\displaystyle r_{|p|}^{(\alpha)}(\xi=1) = K^{(\alpha)}(L^{-\frac{\alpha}{2}})_{|p|} = K^{(\alpha)} (-1)^p \frac{(-\alpha)!}{(\frac{-\alpha}{2}+p)!(\frac{-\alpha}{2}-p)!}  \\ \\
\displaystyle \hspace{1cm} = \frac{\alpha !}{\frac{\alpha}{2}!\frac{\alpha}{2}!} (-1)^p \frac{(-\alpha)!}{(\frac{-\alpha}{2}+p)!(\frac{-\alpha}{2}-p)!} \hspace{0.5cm} > 0 ,\hspace{1cm} 0 < \alpha < 1
\end{array}
\end{equation}
which indeed is consistent with (\ref{spetraleval}) when replacing there $\alpha \rightarrow -\alpha$. For numerical evaluations and to obtain the asymptotic behavior
the following equivalent representation  of (\ref{fundmat3}) is useful\footnote{Which is obtained
by applying Euler's reflection formula in (\ref{fundmat3}) and is also obtained from (\ref{spetraleval}) by replacing $\alpha\rightarrow -\alpha$.}
\begin{equation}
 \label{euler}
  r_{|p|}^{(\alpha)}(\xi=1) = K^{(\alpha)}\frac{(-\alpha)!}{\pi}\sin\left(\frac{\pi\alpha}{2}\right) \, \frac{(|p|+\frac{\alpha}{2}-1)!}{(|p|-\frac{\alpha}{2})!} ,\hspace{1cm} 0<\alpha <1 .
\end{equation}
Let us consider the asymptotic behavior for $|p|>>1$. Since for $\beta >>1$ we have the asymptotics
$\frac{(\beta +a)!}{(\beta+b)!} \sim \beta^{a-b}$. So (\ref{euler}) takes for $|p| >>1$ the asymptotic behavior
\begin{equation}
 \label{rieszpot}
 r_{|p|>>1}^{(\alpha)}(\xi=1) \approx K^{(\alpha)}\frac{(-\alpha)!}{\pi}\sin\left(\frac{\pi\alpha}{2}\right) \, \frac{1}{|p|^{1-\alpha}}   ,\hspace{1cm} 0 < \alpha < 1
\end{equation}
which is
evanescent at infinity. Expression (\ref{rieszpot}) coincides with the inverse kernel of the fractional Laplacian, the {\it Riesz potential}
$(-\frac{d^2}{dp^2})^{-\frac{\alpha}{2}}\delta(p)$ [where $\delta(..)$ denotes Dirac's $\delta$-function] which is in the present case of the 1D infinite space \cite{riesz1949}.
We notice that (\ref{fundmat3}) $ r_{|p-q|}^{(\alpha)}(\xi=1)$ also is a T\"oplitz matrix.

With these results we obtain the probability $F^{(\alpha)}_{|p-q|}$ of ever passage (\ref{fractionaleverret}) for the infinite ring in closed form.
Assuming that the walker starts at node $0$, the probability that the walker ever reaches a node $p$, i.e. a node with distance $|p|$ from the departure node, is obtained as
\begin{equation}
\label{evernodep}
F^{(\alpha)}_{|p|} = \frac{r_{|p|}^{(\alpha)}(\xi=1)-\delta_{p0}}{r_{|0|}^{(\alpha)}(\xi=1)}
\end{equation}
which becomes with (\ref{fundmat3}) an explicit expression.

\begin{figure*}[!t]
\begin{center}
\includegraphics[width=0.55\textwidth]{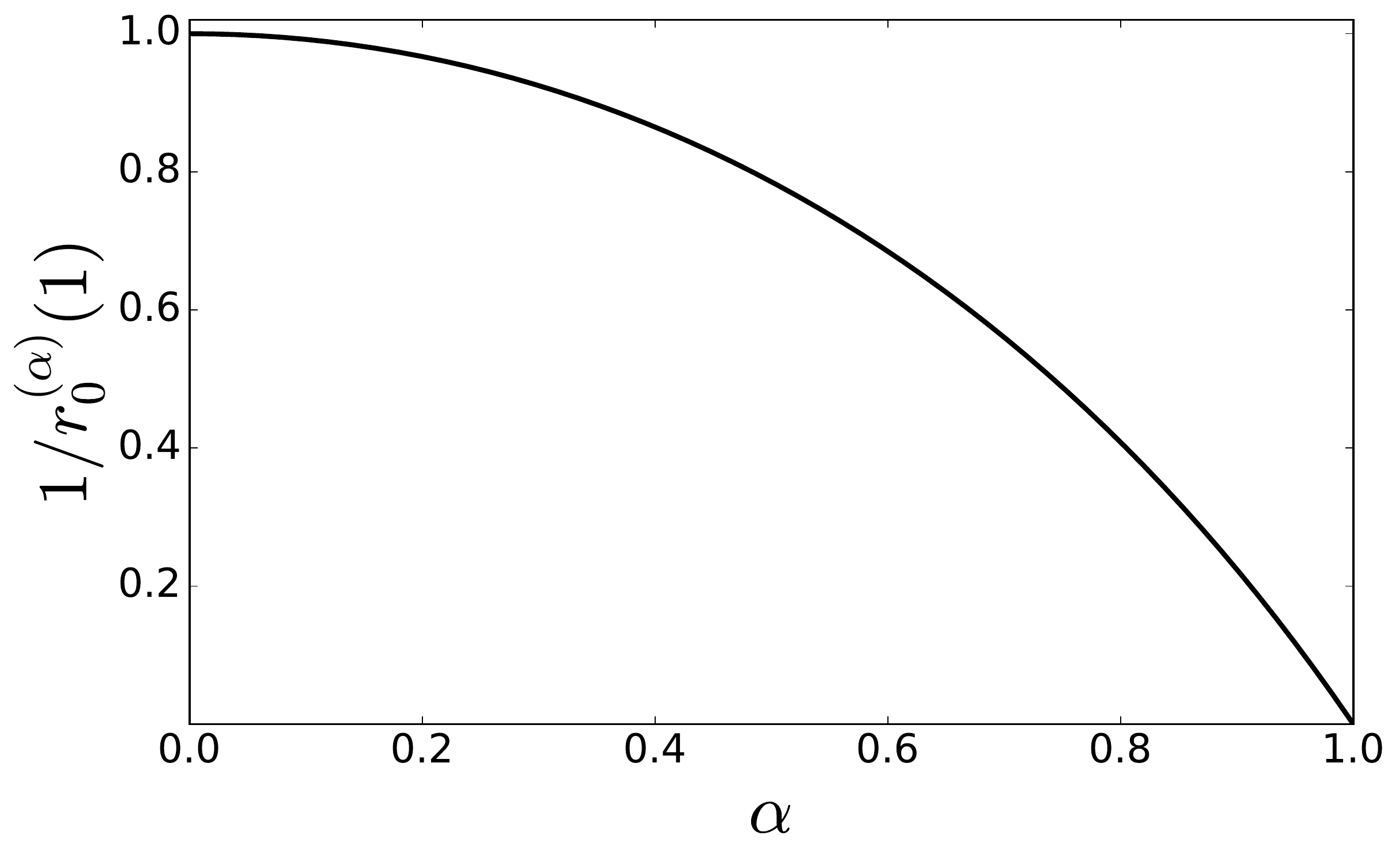}
\end{center}
\vspace{-5mm}
\caption{\label{Figure2} Escape probability $(r_0^{(\alpha)}(1))^{-1}$ versus $\alpha$ in the transient regime  $0<\alpha<1$.
In the limiting case $\alpha\rightarrow 0+$ indicates the limit of extreme transience where the walker is sure to escape.
$\alpha=1-0$ is the recurrent limit where the escape probability is vanishing and the walker is sure to return. }
\end{figure*}

Let us now analyze probabilities of ever return to the departure node (recurrence) and escape probabilities. Accounting for (\ref{fundmat3}) for $p=0$
yields\footnote{Where we denote $K^{(-\alpha)} =  \frac{(-\alpha)!}{(-\frac{\alpha}{2})!(-\frac{\alpha}{2})!}$.}
\begin{equation}
\label{diagelementrmatrix}
 r_0^{(\alpha)}(1) = K^{(\alpha)}K^{(-\alpha)}= \frac{\alpha!}{\frac{\alpha}{2}!\frac{\alpha}{2}!}  \frac{(-\alpha!)}{(\frac{-\alpha}{2})!(\frac{-\alpha}{2})!} =
 \frac{\Gamma(1+\alpha)\Gamma(1-\alpha)}{\Gamma^2(1+\frac{\alpha}{2})\Gamma^2(1-\frac{\alpha}{2})}
 \hspace{1cm} 0<\alpha<1
\end{equation}
which are well defined expressions within the transient regime $0<\alpha<1$ with the escape probability (probability of never return to the departure node) $\frac{1}{r^{(\alpha)}(1)}$. From above general expression (\ref{relation})
we observe that for $\alpha\rightarrow 0$ the escape probability $\frac{1}{r^{(\alpha)}(1)} \rightarrow 1-0$ which is recovered by (\ref{diagelementrmatrix}):
This limit of extreme transience (sure escape of the walker at $\alpha\rightarrow 0$) is in accordance with above obtained general relation (\ref{extreme}).
\\[2mm]
In Figure \ref{Figure2} is plotted the escape probability $(r^{(\alpha)}(1))^{-1}$ of the explicit
expression (\ref{diagelementrmatrix}) for the transient regime $0<\alpha<1$ for 1D ($d=1$).
We further notice that in the limit $\alpha\rightarrow 1$ the relation (\ref{diagelementrmatrix}),
due to $\Gamma(1-\alpha) \rightarrow\infty$, tends to infinity and as a consequence the escape probability is vanishing.
This is consistent with above recurrence theorem as $\alpha=1$ constitutes for $d=1$ the limit of recurrence. We emphasize that expression (\ref{diagelementrmatrix})
exists only in the transient interval $0<\alpha<1$ of case (i).
\newline\newline
Let us return to the interpretation below relation (\ref{kemenycosnat}): Figure \ref{Figure2} shows that the escape probability $(r_0^{(\alpha)}(1))^{-1} \sim N/{\cal K}_e^{(\alpha)}$
being the infinite lattice limit of the renormalized
inverse Kemeny constant (inverse renormalized global MFPT) $({\cal K}_e^{(\alpha)}/N)^{-1}$ being a measure how fast the walker visits a randomly selected node different from the departure node.
The more transient the FRW is for small $\alpha$, the more fast in the infinite network the walk necessarily is.

In view of our above mentioned interpretation of the Green's function, namely that $r_0^{(\alpha)}(1)$ counts the average number of time steps (the MRT) the walker is present in the departure node $p=0$ (i.e. $r_0^{(\alpha)}(1)-1$ indicates the average number of {\it returns} to the departure node) we observe in Figure \ref{Figure2} that in the extreme transient limit
$\alpha\rightarrow 0+$ as $r_0^{(\alpha\rightarrow 0+)}(1)=1$ the average number of returns is vanishing.

Now, let us discuss the ever passage probabilities (\ref{evernodep}) for $p\neq 0$, i.e. for nodes different as the departure node. Then (\ref{evernodep}) assumes the form
\begin{equation}
 \label{everpassagenonzeronode}
 F^{(\alpha)}_{|p|} = \frac{r_{|p|}^{(\alpha)}(\xi=1)}{r_{|0|}^{(\alpha)}(\xi=1)}=\frac{(L^{-\frac{\alpha}{2}})_{|p|}}{(L^{-\frac{\alpha}{2}})_{0}}  ,\hspace{1cm} p\neq 0
\end{equation}
which takes with (\ref{fundmat3})
\begin{equation}
\label{explcitfirstpass}
F^{(\alpha)}_{|p|} = (-1)^p\frac{(-\frac{\alpha}{2})!(-\frac{\alpha}{2})!}{(-\frac{\alpha}{2}+p)!(-\frac{\alpha}{2}-p)!}=
(-1)^p\frac{\Gamma^2(1-\frac{\alpha}{2})}
{\Gamma(1-\frac{\alpha}{2}+p)\Gamma(1-\frac{\alpha}{2}-p)}    ,\hspace{1cm} 0<\alpha<1
\end{equation}
with $p\neq 0$ where $|p|$ indicates the distance of the departure node with $0<\alpha<1$.
In view of the initial representation (\ref{fundmat}) the property $0< F^{(\alpha)}_{|p|}<1$ reflecting the probability interpretation is verified in the appendix \ref{append}.

\begin{figure*}[!t]
\begin{center}
\includegraphics*[width=0.6\textwidth]{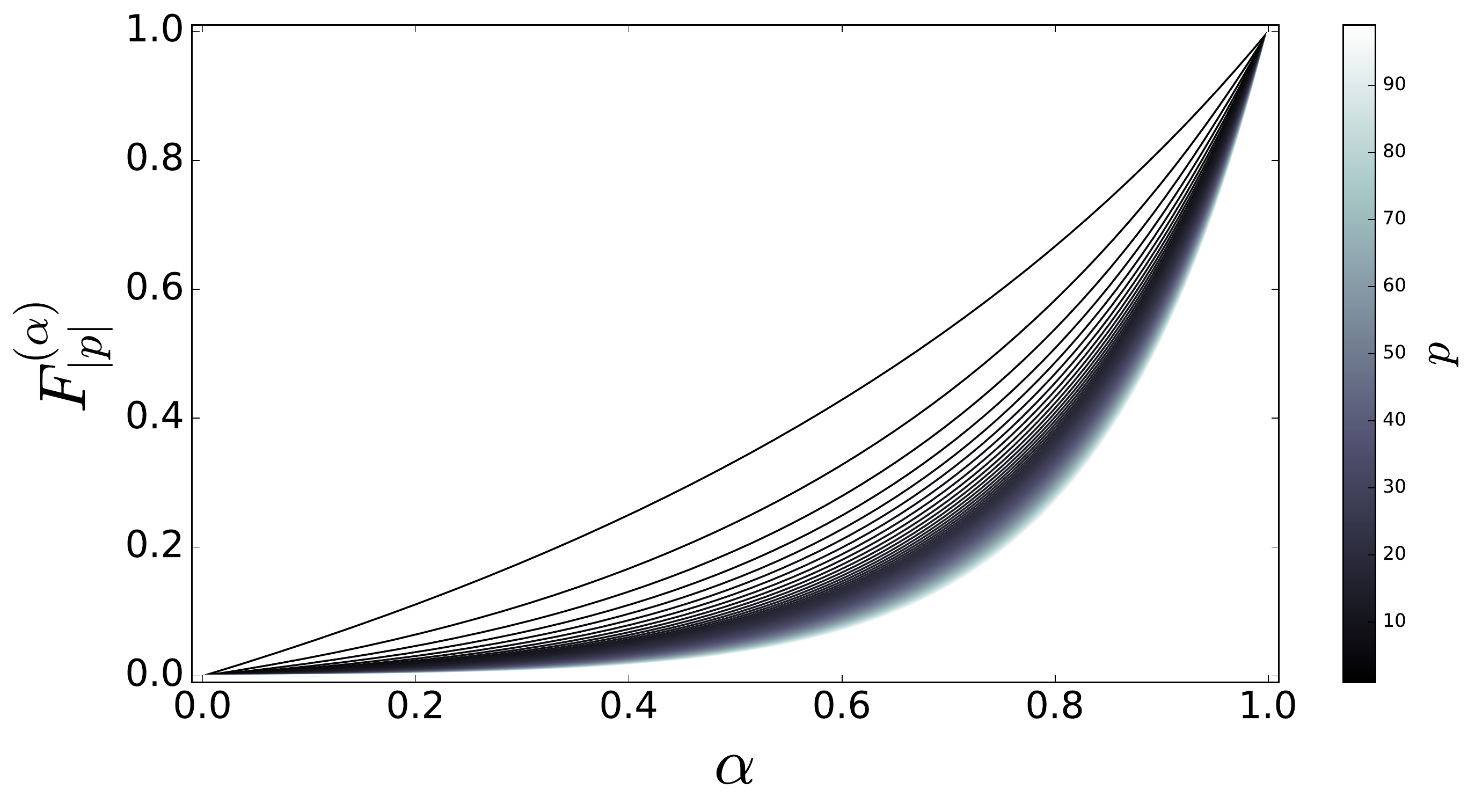}
\end{center}
\vspace{-8mm}
\caption{\label{Figure3} Ever passage probabilities $F^{(\alpha)}_{|p|}$ as a function of $\alpha$ for different values of $p$. The results are obtained by numerical evaluation of  relation (\ref{explcitfirstpass}), each curve is  depicted with colors that allow to identify the respective value of $p$ codified in the colorbar. }
\end{figure*}

\begin{figure*}[!h]
\begin{center}
\includegraphics*[width=0.65\textwidth]{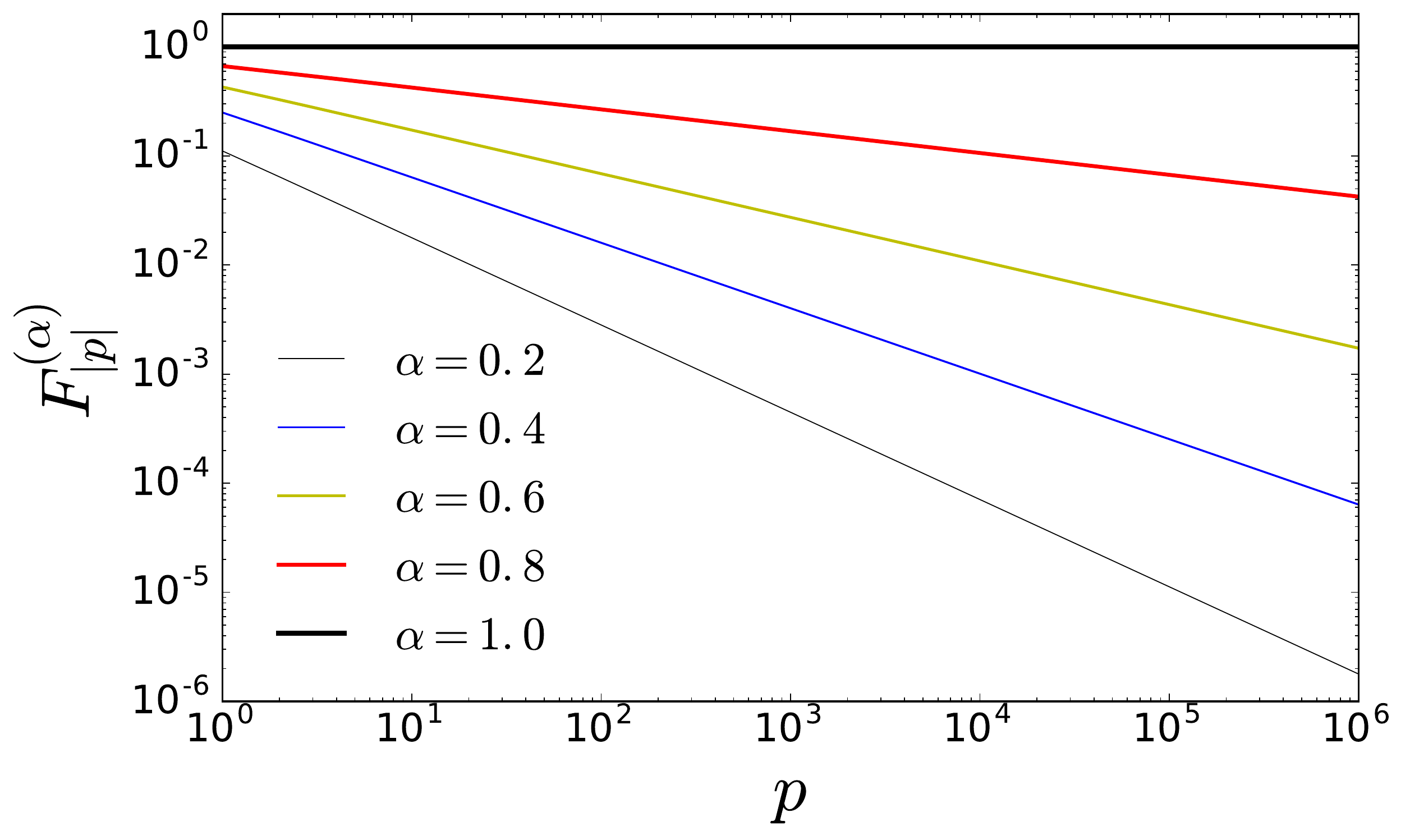}
\end{center}
\vspace{-8mm}
\caption{\label{Figure4} Ever passage probabilities $F^{(\alpha)}_{|p|}$ as a function of $p$ for different values of $\alpha$. The results are obtained by numerical evaluation of  relation (\ref{explcitfirstpass}).
We observe straight lines in the logarithmic scale for $p>>1$, result that agrees with the asymptotic relation $F^{(\alpha)}_{|p|}\sim |p|^{\alpha-1}$. }
\end{figure*}

In Figures \ref{Figure3}, \ref{Figure4}, the ever passage probabilities $F^{(\alpha)}_{|p|}$  for the transient regime $0<\alpha<1$ are drawn.
In Figure \ref{Figure3} we depict $F^{(\alpha)}_{|p|}$ as a function of $\alpha$ for different nodes $|p|=1,\ldots,100$.
We observe what we see analytically for $\alpha\rightarrow 0$ in (\ref{euler}), that $\lim_{\alpha\rightarrow 0+}F^{(\alpha)}_{|p|}=0$. This holds for all $p$ including
$p=0$ (see Figure  \ref{Figure2}) where $\lim_{\alpha\rightarrow 0} F^{(\alpha)}_{|0|}= \lim_{\alpha\rightarrow 0} (1-(r^{(\alpha)}(1))^{-1})=0$ as $r^{(\alpha\rightarrow 0+)}(1)= 1$ as
demonstrated by above general relation (\ref{extreme}). On the other hand we observe in Figure \ref{Figure3} and also in Figure \ref{Figure2}, that as we approach the limit $\alpha=1$ of
recurrence that the ever passage probability for all nodes approaches the value
$\lim_{\alpha\rightarrow 1-0} F^{(\alpha)}_{|p|} = 1$
of sure ever passage which also follows directly from relations (\ref{euler}) together with (\ref{everpassagenonzeronode}) and (\ref{everpassageproduct}).
Finally, in Figure \ref{Figure4} we plot $F^{(\alpha)}_{|p|}$
as a function of $p$ for the values of $\alpha=0.2,\, 0.4,\, 0.6,\, 0.8,\, 1.0$.
We notice that $F^{(\alpha)}_{|p|}$ for a fix $\alpha$ is
monotonously decreasing with increasing $|p|$ (the result is a power-law relation approaching to zero as a
Riesz potential $\sim |p|^{\alpha-1}$ for $|p|\rightarrow \infty$, see asymptotic relation (\ref{rieszpot}) and appendix \ref{appendb}). In the appendix \ref{append}
we give a short proof that $F^{(\alpha)}_{|p+1|} < F^{(\alpha)}_{|p|}$
which also can be seen in Figures \ref{Figure3}, \ref{Figure4}.

\section{Conclusions}

In this paper we have analyzed FRWs on regular networks, especially $d$-dimensional simple cubic lattices in the framework of Markovian processes. The FRW generalizes the Polya walk by replacing
the Laplacian matrix $L$ by a fractional power $L^{\frac{\alpha}{2}}$ with $0<\alpha \leq 2$ allowing within $0<\alpha<2$ long range moves where in sufficiently large networks the
probability of occurence of long range steps decays
as an inverse power law heavy tailed (L\'evy-) distribution $W^{(\alpha)}_{|\vec{p}-\vec{q}|} \sim |{\vec p}-\vec{q}|^{-\alpha-d}$ of the form of the Fractional Laplacian operator
kernel (Riesz fractional derivative) of the $d$-dimensional infinite space. This property is a landmark of the emergence of L\'evy flights in sufficiently `large' lattices
\cite{riascos-fracdyn,riascos-fracdiff,michelitsch-riascos2017}. [For a brief demonstration, see appendix \ref{appendb}, relation (\ref{transitionfractionaldinfiniteassymp})].
The tendency in the FRW to perform long-range steps increases as $\alpha$ decreases. As a consequence the speed of the walk,
and finally - in the infinite lattice- the escape probability
increases the smaller $\alpha$ [measured by the renormalized inverse Kemeny constant $N({\cal K}_e^{(\alpha)})^{-1}$] which includes the limiting cases of
extreme transience at $\alpha=0+$ with sure escape (Depicted in Figure \ref{Figure2} for the infinite ring).

We established for $d$-dimensional infinite lattices a generalization of Polya's recurrence theorem to Fractional Random Walks
and random walks having the same (heavy tailed) inverse power law characteristics which include L\'evy flights: FRWs are transient for lattice
dimensions $d>\alpha$ and recurrent for $d\leq \alpha$ where $\alpha$ always
is restricted to
$0<\alpha\leq 2$. For the strongly
transient regime $0<\alpha<1$ where the FRW is transient for all lattice dimensions we have obtained for the infinite ring ($d=1$) closed form expressions for the fractional lattice Green's function
containing the complete statistical information such as the ever passage and escape probabilities.
In the limiting case of extreme transience ($\alpha\rightarrow 0+$) the walker is sure to escape. In contrast in the recurrent limit $\alpha\rightarrow 1-0$ the escape probability approaches zero
(Figure \ref{Figure2}).
In the extreme transient case $\alpha\rightarrow 0+$ the ever passage probability for any fixed node $p$ approaches zero. This can be understood
by overleaping of nodes (especially close to the departure node) due to frequent long range steps. The oppositive effect takes place
when approaching the recurrent limit $\alpha\rightarrow 1-0$: The walker is sufficiently slow with less long-range steps thus nodes located `not too far away from the departure node' become sure to be visited. {\it As a consequence a searched target on such nodes becomes sure to be found} (Figures. \ref{Figure3},\ref{Figure4}).

The overall significance of the Fractional Random Walk is the interplay of a discrete random motion on a well defined set (network) and L\'evy motions with emergence of asymptotically scale free nonlocality
$ \sim |{\vec p}-\vec{q}|^{-\alpha-d}$ for the probabilities of large steps on sufficiently large networks. This scale free nonlocality of the FRW is  independent of the details of the generating Laplacian $L$ and therefore is an {\it universal} small world feature
depending only on the dimension $d$ of the lattice and the scaling index $\alpha$.

In view of its universal properties, the FRW deserves further investigations, especially in the context of dynamical processes in complex networks (information exchange in complex systems).
This includes also the analysis of recurrence behavior of {\it L\'evy walks on networks}, i.e. when the velocity of the walker is finite \cite{klafter,barkai}.

The remarkably rich dynamics behind FRWs should be further analyzed, for instance when passing from
random walks to random flights where the latter take place in the limit of a continuous distribution of nodes in continuous spaces.
Especially first passage events of FRWs deserve further
attention since they play a key role in the understanding of chemical reactions and chaotic turbulent motions, population dynamics and in a vast number of interdisciplinary problems.

\section{Acknowledgements}
We thank Eli Barkai for valuable comments pointing out different recurrence features of L\'evy flights and L\'evy walks.

\begin{appendix}
\section{Appendix 1}
\label{append}

We analyze some necessary properties of expression (\ref{explcitfirstpass}): First we have to prove that $0 \leq F^{(\alpha)}_{|p|} \leq 1$ allowing probability interpretation:

\begin{equation}
\label{everpassage}
 F^{(\alpha)}_{|p|} = (-1)^p\frac{(-\frac{\alpha}{2})!(-\frac{\alpha}{2})!}{(-\frac{\alpha}{2}+p)!(-\frac{\alpha}{2}-p)!}=
(-1)^p\frac{\Gamma^2(1-\frac{\alpha}{2})}
{\Gamma(1-\frac{\alpha}{2}+p)\Gamma(1-\frac{\alpha}{2}-p)} ,\hspace{1cm} (p\neq 0),\hspace{1cm} 0 <\alpha < 1 .
\end{equation}
Using the property $(\zeta + p)! = \zeta ! \prod_{s=1}^p(\zeta +s)$ and by setting $\zeta= -\frac{\alpha}{2}-p$ gives

$\frac{(-\frac{\alpha}{2})!}{( -\frac{\alpha}{2}-p)!}=\prod_{s=1}^p(-\frac{\alpha}{2}-p +s) =(-1)^p\prod_{s=0}^{p-1}(\frac{\alpha}{2}+s)$
and further $\frac{(-\frac{\alpha}{2})!}{(-\frac{\alpha}{2}+p)!} = \prod_{s=0}^{p-1}\frac{1}{(1-\frac{\alpha}{2}+s)}$ thus we can write for
(\ref{everpassage}) the product representation
\begin{equation}
\label{everpassageproduct}
 F^{(\alpha)}_{|p|} = \prod_{s=0}^{p-1}\frac{(\frac{\alpha}{2}+s)}{(1-\frac{\alpha}{2}+s)} ,\hspace{1cm} (p\neq 0),\hspace{1cm}  0 <\alpha < 1 .
\end{equation}
The observation is that as $1-\frac{\alpha}{2} > 0$ all factors are positive thus $F^{(\alpha)}_{|p|} >0 $. Now because of $0<\alpha<1$ we can put
$\frac{\alpha}{2}= \frac{1}{2}-\epsilon$ ($\epsilon >0$) thus
\begin{equation}
\label{everpassageproductepsilon}
0<  F^{(\alpha)}_{|p|} = \prod_{s=0}^{p-1}\frac{(\frac{1}{2}+s-\epsilon)}{(\frac{1}{2}+s+\epsilon)} <1 ,\hspace{1cm} (p\neq 0), \hspace{0.5cm} 0 <\alpha < 1 .
\end{equation}
It follows that $0< F^{(\alpha)}_{|p|} < 1$ as each factor fulfills $0< \frac{(\frac{1}{2}+s-\epsilon)}{(\frac{1}{2}+s+\epsilon)} < 1$ ($\epsilon=\frac{1}{2}-\frac{\alpha}{2} >0 $).
Especially we observe that
\begin{equation}
 \label{decrease}
  F^{(\alpha)}_{|p+1|} = \frac{(\frac{1}{2}+p-\epsilon)}{(\frac{1}{2}+p+\epsilon)}  F^{(\alpha)}_{|p|}
\end{equation}
and hence $F^{(\alpha)}_{|p+1|} < F^{(\alpha)}_{|p|}$ that is the ever passage probability decays monotonously when the distance $|p|$ from the departure node increases.
We hence have proved that $0< F^{(\alpha)}_{|p|}<1$ for $0<\alpha<1$ as a necessary condition allowing (ever passage) probability interpretation. We further observe in
$F^{(\alpha=0+)}_{|p|} = 0$ (extreme transience) and $F^{(\alpha=1-0)}_{|p|} = 1$ (recurrence), see Figure \ref{Figure2}.

\section{Appendix 2}
\label{appendb}
Here our goal is to briefly demonstrate the
asymptotic behavior for the transition matrix (\ref{transitionfractional}) for $|\vec{p}-\vec{q}|>>1$ and $N_j\rightarrow\infty$ $j=1,..,d$ in the fractional interval $0<\alpha<2$
as a landmark for the emergence of L\'evy flights in sufficiently large $d$-dimensional lattices. The spectral representation of the transition matrix is
\begin{equation}
 \label{transitionfractionaldinfinite}
 {\cal W}^{(\alpha)}(\vec{p}-\vec{q}) = \sum_{\vec{\ell}} \lambda_{\vec{\ell}}^{(\alpha)} \frac{e^{i{(\vec{p}-\vec{q})\cdot\vec{\kappa}_{\vec{\ell}}}}}{N}
 ,\hspace{0.5cm} \lambda_{\vec{\ell}}^{(\alpha)} = 1- \frac{\mu^{\frac{\alpha}{2}}(\vec{\kappa}_{\vec{\ell}})}{K^{(\alpha)}} ,\hspace{1cm}  0<\alpha \leq 2 .
\end{equation}
Consider now the
probability that the walker makes a long-range move of $|\vec{p}-\vec{q}|>>1$.
Using $(\mu(\vec{\kappa}))^{\frac{\alpha}{2}} \sim |\vec{\kappa}|^{\alpha} $ for $|\vec{\kappa}|\rightarrow 0$ the principal contribution to
the fractional adjacency matrix elements writes
\begin{equation}
 \label{transitionfractionaldinfiniteassymp}
A^{(\alpha)}(\vec{p}-\vec{q})) = K^{(\alpha)} {\cal W}^{(\alpha)}(\vec{p}-\vec{q}) \approx - \frac{1}{(2\pi)^d}\int |\vec{\kappa}|^{\alpha} e^{i{(\vec{p}-\vec{q})\cdot\vec{\kappa}}}
=-\left(-\Delta_{(\vec{p}-\vec{q})}\right)^{\frac{\alpha}{2}}\delta^d(\vec{p}-\vec{q}) =  \frac{C_{\alpha,d}}{|\vec{p}-\vec{q}|^{d+\alpha}}
 \end{equation}
with the positive constant $C_{\alpha,d}=\frac{2^{\alpha-1}\alpha\Gamma(\frac{\alpha+d}{2})}{\pi^{\frac{d}{2}}\Gamma(1-\frac{\alpha}{2})} $ \cite{michel} where the latter inverse power law kernel holds for $0<\alpha<2$ ($\alpha \neq 2$). The transition matrix
(probability of a long range jump of distance
$|(\vec{p}-\vec{q})|$) thus scales in the fractional interval $0<\alpha<2$ as an inverse power law
${\cal W}^{(\alpha)}(\vec{p}-\vec{q}) \sim |\vec{p}-\vec{q}|^{-d-\alpha}$
having the form of the kernel of the fractional Laplacian operator (Riesz fractional derivative)
in the $d$-dimensional infinite space \cite{michel,michelitsch-riascos2017}. Returning to the master equation (\ref{timemaster}) the time evolution of the occupation probabilities
with (\ref{transitionfractionaldinfiniteassymp}) is for $|\vec{p}-\vec{q}| >>1$ asymptotically described by (with $P_{t+1}(\vec{p}-\vec{q})-P_t(\vec{p}-\vec{q}) \approx \frac{P_{t}(\vec{p})}{dt}$)
\begin{equation}
 \label{levyflight}
 \frac{P_{t}(\vec{p}-\vec{q})}{dt} \approx -\frac{(L^{\frac{\alpha}{2}})_{\vec{p}-\vec{q}}}{K^{(\alpha)}} \approx
 -\frac{1}{K^{(\alpha)}}\left(-\Delta_{(\vec{p}-\vec{q})}\right)^{\frac{\alpha}{2}}\delta^d(\vec{p}-\vec{q})
\end{equation}
which is the evolution equation of a (time-continuous) L\'evy flight in $d$-dimensional infinite space with L\'evy index $0<\alpha<2$ where $\delta^d(\vec{p}-\vec{q})$
denotes the $d$-dimensional Dirac's $\delta$-function \cite{michelitsch-riascos2017}. Asymptotic relation (\ref{levyflight}) recovers for $\alpha=2$ the conventional diffusion equation indicating the Brownian nature of the Polya walk.

By a similar consideration the asymptotic representation of the fractional lattice Green's function (\ref{rpqalpha}) of the $d$-dimensional infinite lattice for
$|\vec{p}-\vec{q}| >>1$ for the transient regime $d-\alpha >0$ is obtained in
Riesz potential form (see also \cite{sato}, pp. 261)
\begin{equation}
 \label{rieszddim}
\frac{1}{K^{(\alpha)}} r^{(\alpha)}(\vec{p}-\vec{q}) \approx \left(-\Delta_{(\vec{p}-\vec{q})}\right)^{-\frac{\alpha}{2}}\delta^d(\vec{p}-\vec{q})= -\frac{C_{-\alpha,d}}{ |\vec{p}-\vec{q}|^{d-\alpha}} >0
\end{equation}
where this expression formally is obtained when replacing $\alpha\rightarrow -\alpha$ (and adding multiplyer $(-1)$) in (\ref{transitionfractionaldinfiniteassymp}). It is worth noticing that the constant $-C_{-\alpha,d} =
2^{-\alpha}\pi^{-\frac{d}{2}}\frac{\Gamma(\frac{d-\alpha}{2})}{\Gamma(\frac{\alpha}{2})} > 0$
occuring in (\ref{rieszddim}) is positive.
It is further worthy to mention that asymptotic relation (\ref{rieszddim}) holds also for the transient Polya walks $d>\alpha=2$.  For a Polya walk on a $3$-dimensional  lattice (\ref{rieszddim}) takes the representation  of a Newtonian potential $ \frac{1}{4\pi |\vec{p}-\vec{q}|} = \left(-\Delta_{(\vec{p}-\vec{q})}\right)^{-1}\delta^3(\vec{p}-\vec{q})$ as a landmark of the Brownian nature of the Polya walk.

Returning to the interpretation of the components of the lattice Green's function (\ref{rpqalpha}) indicating
the average number of visits of nodes (the MRT): Let us briefly consider the MRT of the walker in a large sphere in the $d$ dimensional space of radius $R>>1$ (sufficiently large that the L\'evy flight characteristics of the FRW emerges).
The walker is assumed to depart in the origin of the sphere: Integrating (\ref{rieszddim}) over this sphere yields for the
MRT a behavior $\sim R^{\alpha}$  independent of the lattice dimension $d$ for the transient regime $d>\alpha$ whereas
the Green's function (\ref{rpqalpha}) and hence the MRT diverges in the recurrent
regime $d\leq \alpha$.
 This observation coincides with the results obtained in \cite{barkai} (Eq. (32) in that paper) for the MRT in the transient regime $0<\alpha<1$ for a L\'evy flyer in the one-dimensional space $d=1$ (and for $d=1\leq \alpha \leq 2$ the MRT diverges in the recurrent regime \cite{barkai}).

(\ref{rieszddim}) remains uniquely positive allowing probability interpretation of the ever passage probabilities:
The probability that the walker in the transient regime ($d>\alpha$ , $0<\alpha\leq 2$) ever reaches a far distant node $|\vec{p}-\vec{q}| >>1$
from the departure node is with (\ref{rieszddim}) and (\ref{fractionaleverret}) given by the inverse power law
\begin{equation}
 \label{fractionalever}
  F^{(\alpha)}_{\vec{p}-\vec{q}}(\xi=1) = \frac{r^{(\alpha)}_{\vec{p}-\vec{q}}(1)}{r^{(\alpha)}_{\vec{0}}(1)} \approx \frac{D_{-\alpha,d}}{ |\vec{p}-\vec{q}|^{d-\alpha}} ,\hspace{1cm} d>\alpha ,\hspace{0.5cm} 0<\alpha\leq 2
\end{equation}
of a Riesz potential type \cite{riesz1949} where the constant $D_{-\alpha,d} =\frac{K^{(\alpha)}(-C_{-\alpha,d})}{r^{(\alpha)}_{\vec{0}}(1)} >0$ is uniquely positive. We further mention that also the asymptotic representations (\ref{rieszddim}) and (\ref{fractionalever})
are {\it universal} as they do not depend on the details of the generating Laplacian matrix $L$.
For the infinite ring $d=1$ (\ref{fractionalever}) takes the Riesz potential power law asymptotics $\sim p^{\alpha-1}$ of explicit expression (\ref{everpassagenonzeronode}).

\end{appendix}


\begin{thebibliography}{1}

\bibitem{newmann} M. E. J. Newman, Networks: An Introduction. Oxford, England: Oxford University Press, 2010.
\bibitem{albert2002} R. Albert and A.-L. Barab\`asi, ``Statistical mechanics of complex networks,'' Rev. Mod. Phys. 74, 47-97 (2002).
\bibitem{NohRieger} J.D. Noh, H. Rieger, Random Walks on Complex Networks, Phys. Rev. Lett. 92, no. 11, 118701 (2004).
\bibitem{gonzales} B. Gon{\c c}alves, N. Perra, and A. Vespignani, “Modeling users’ activity on Twitter networks: Validation of Dunbar’s number,” PLoS ONE, vol. 6, p. e22656, 08 2011.
\bibitem{rathkiewics} J. Ratkiewicz, S. fto, A. Flammini, F. Menczer, and A. Vespignani, “Characterizing and modeling
the dynamics of online popularity,” Phys. Rev. Lett., vol. 105, p. 158701, Oct 2010.
\bibitem{riascos12} A. P. Riascos, J. L. Mateos, Long-range navigation on complex networks using L\'evy random walks, Phys. Rev. E 86, 056110 (2012)

\bibitem{getoor} R.K. Getoor, First passage times for symmetric stable processes in space, Trans. Amer. Math. Soc. 101 (1961), 75-90.

\bibitem{blumenthal-getoor} R.M. Blumenthal, R.K. Getoor, D.B. Ray, On the distribution of first hits for the symmetric stable processes,  Trans. Amer. Math. Soc. 99 (1961), 540-554.


\bibitem{metzler1} R. Metzler, J. Klafter, The random walk’s guide to anomalous diffusion: a fractional dynamics approach Phys. Rep. 339 1–77 (2000).

\bibitem{klafter} V. Zaburdaev, S. Denisov, J. Klafter,  L\'evy walks, Rev. Mod. Phys. 87, 483 (2015).   

\bibitem{barkai} B. Dybiec, E.  Gudowska-Nowak, E. Barkai, A.A. Dubkov, L\'evy flights versus L\'evy walks in bounded domains, Phys.  Rev. E 95, 052102 (2017). 

 \bibitem{Dybiec-Nowac} B. Dybiec, E. Gudowska-Nowak, A. Chechkin, To hit or to pass it over—remarkable
transient behavior of first arrivals and passages for L\'evy flights in finite domains, 
J. Phys. A: Math. Theor. 49 504001 (2016).

\bibitem{ferraro} M. Ferraro and L. Zaninetti, Mean number of visits to sites in Levy flights,
Phys. Rev. E 73, 057102 (2006). 

\bibitem{sato} K. Sato, L\'evy processes and Infinitely Divisible Distributions, Cambridge University Press, 1999, (Cambridge Studies in Advanced Mathematics 68).

\bibitem{tarasov} V.E. Tarasov, Lattice fractional calculus Appl. Math. Comput. 257 1233 (2015).

\bibitem{ortiguera} M.D. Ortigueira, Riesz potential operators and inverses via fractional centered derivatives
Int. J. Math. Math. Sci. 48391 112 (2006).

\bibitem{zhang} Z. Zhang, A. Julaiti, B. Hou, H. Zhang, G. Chen, Mean first passage time for random walks on undirected networks, Eur. Phys. J. B 84, 691 (2011).

\bibitem{polya} G. P\'olya, \"Uber eine Aufgabe der Wahrscheinlichkeitsrechnung betreffend die Irrfahrt im Straßennetz, Mathematische Annalen 83 (1921), 149-160.
\bibitem{montroll} E.W. Montroll, Random Walks in Multidimensional Spaces, Especially on Periodic Lattices. J. SIAM 4, 241-260, (1956).
\bibitem{montroll-weiss} E.W. Montroll, G.H. Weiss, Random Walks on Lattices. II, J. Math. Phys. Vol 6, No. 2; 167-181 (1965). doi:10.1063/1.1704269

\bibitem{hudges} B.D. Hudges, Random walks and random environments, Oxford University press Inc New York 1995, ISBN 0 19 853788 3.

\bibitem{Watts} D.J. Watts, S.H. Strogatz, Nature (London) 393, 440 (1998).
\bibitem{dorogotsev} S. N. Dorogovtsev, A. V. Goltsev, Critical phenomena in complex networks, Rev. Mod. Phys. 80, 1275-1335 (2008).

\bibitem{Erdos} P. Erd\"os, A. R\'enyi, “On random graphs, I,” Publicationes Mathematicae (Debrecen), vol. 6, pp. 290-297, 1959.

\bibitem{miegem} P. V. Mieghem, Graph Spectra for Complex Networks. New York, NY, USA: Cambridge University Press, 2011.

\bibitem{doyle} P. G. Doyle, J. Laurie Snell, (1984), Random Walks and Electric Networks. Carus Mathematical Monographs 22.
\bibitem{kemeny} John G. Kemeny, J. Laurie Snell, Finite Markov Chains, Springer Verlag, New York, Berlin, Tokyo, 1976.


\bibitem{riascos-fracdyn} A.P. Riascos, J.L. Mateos, Fractional dynamics on networks:
Emergence of anomalous diffusion and L\'evy flights, Phys. Rev. E 90, 032809 (2014). (arxiv:1506.06167).

\bibitem{riascos-fracdiff} A.P. Riascos, J.L. Mateos, Fractional diffusion on circulant networks: emergence of a dynamical small world, J. Stat. Mech. (2015) P07015.


\bibitem{michelitsch-riascos2017} T.M. Michelitsch, B. Collet, A.F Nowakowski, A.P. Riascos, A.F. Nowakowski, F.C.G.A. Nicolleau, Fractional random walk lattice dynamics,
J. Phys. A: Math. Theor. 50 (2017) 055003 . doi:10.1088/1751-8121/aa5173 .


\bibitem{michelJphysA} T.M. Michelitsch, B. Collet, A.F Nowakowski, F.C.G.A. Nicolleau,
Fractional Laplacian matrix on the finite periodic linear chain and its periodic Riesz fractional derivative continuum limit ,
J. Phys. A: Math. Theor. 48 295202  (2015). (arXiv:1412.5904).

\bibitem{michelchaos} T.M. Michelitsch, B. Collet, A.F. Nowakowski, F.C.G.A.
Nicolleau, Lattice fractional Laplacian and its continuum limit kernel on the finite cyclic chain, Chaos, Solitons \& Fractals 82, pp. 38-47 (2016).  	
(arXiv:1511.01251)

\bibitem{michelchaos2}
T.M. Michelitsch, B.A. Collet, A.P. Riascos, A.F. Nowakowski, F.C.G.A. Nicolleau, A fractional generalization of the classical lattice dynamics approach,
Chaos, Solitons \& Fractals 92 (2016) 43–50.

\bibitem{Zoia2007} A. Zoia A, A. Rosso, M. Kardar, Fractional Laplacian in bounded domains,
Phys. Rev. E 76, 021116 (2007).

\bibitem{Feller} W. Feller, An Introduction to Probability Theory and its Applications, John Wiley \& Sons Inc. New York, London 1950.


\bibitem{abramo} Abramovitz and Stegun, Handbook of Mathematical Functions, http://people.maths.ox.ac.uk/~macdonald/aands/index.html

\bibitem{gelfand} I. M. Gel'fand and G. E. Shilov , Generalized Functions, AMS Chelsea Publishing, Volume 1, 1964, ISBN-10: 1-4704-2658-7
ISBN-13: 978-1-4704-2658-3.

\bibitem{riesz1949} Riesz, Marcel (1949), "L'int\'egrale de Riemann-Liouville et le probl\`eme de Cauchy", Acta Mathematica, 81: 1–223, doi:10.1007/BF02395016, ISSN 0001-5962, MR 0030102.

\bibitem{michel} T.M. Michelitsch, G.A. Maugin, D. Derogar S and M. Rahman, A regularized representation of the fractional
Laplacian in n dimensions and its relation to Weierstrass–Mandelbrot-type fractal functions
IMA J. Appl. Math. 79 753777 (2à14).


\end{thebibliography}
\end{document}